\documentclass{amsart}
\usepackage{graphicx} % Required for inserting images
\usepackage{amssymb}
\usepackage[dvipsnames]{xcolor}
\usepackage{amsmath}
\usepackage{bbm}
\usepackage{mathptmx}
\usepackage{multirow}
\DeclareMathOperator{\RX}{\mathbb{R}}% real field
\DeclareMathOperator{\XX}{\boldsymbol{X}}
\DeclareMathOperator{\YX}{\boldsymbol{Y}}
\DeclareMathOperator{\SX}{\boldsymbol{S}}

\DeclareMathOperator{\GX}{\mathbb{G}}

\begin{document}
%\title[SMARTA+]{Efficient Removal of Transient DC and Stimulus Artifacts for Adaptive Deep Brain Stimulation and a Temporal Event Localization Analysis}

\title[SMARTA+]{Efficient Artifacts Removal for Adaptive Deep Brain Stimulation and a Temporal Event Localization Analysis}

\author{Tzu-Chi Liu}\thanks{T.-C. Liu: Neuroscience Research Center, Chang Gung Memorial Hospital, Taoyuan, Taiwan; Department of Mathematics, National Taiwan University, Taipei, Taiwan.}
\author{Po-Lin Chen}\thanks{P.-L. Chen: Neuroscience Research Center and Division of Movement Disorders, Department of Neurology, Chang Gung Memorial Hospital, Taoyuan, Taiwan.}
\author{Yi-Chieh Chen}\thanks{Y.-C. Chen: Division of Movement Disorders, Department of Neurology, and College of Medicine, Chang Gung University, Taoyuan, Taiwan.}
\author{Po-Hsun Tu}\thanks{P.-H. Tu: College of Medicine, Chang Gung University, Taoyuan, Taiwan; Department of Neurosurgery, Chang Gung Memorial Hospital at Linkou; School of Medicine, National Tsing Hua University, Hsinchu, Taiwan.}
\author{Chih-Hua Yeh}\thanks{C.-H. Yeh: Department of Neuroradiology, Chang Gung Memorial Hospital, Linkou, Taiwan.}
\author{Mun-Chun Yeap}\thanks{M.-C. Yeap: Department of Neurosurgery, Chang Gung Memorial Hospital at Linkou, Taoyuan, Taiwan.}
\author{Chiung-Chu Chen}\thanks{C.C. Chen, co-corresponding author: neurozoe@gmail.com; affiliated with Neuroscience Research Center, Division of Movement Disorders, and College of Medicine, Chang Gung University, Taoyuan, Taiwan.}
\author{Hau-Tieng Wu}\thanks{H.-T. Wu, co-corresponding author: hauwu@cims.nyu.edu; Courant Institute of Mathematical Sciences, New York University, New York, USA.}

\begin{abstract}
\textbf{Background:} Adaptive deep brain stimulation (aDBS) leverages symptom-related biomarkers to deliver personalized neuromodulation therapy, with the potential to improve treatment efficacy and reduce power consumption compared to conventional DBS. However, stimulation-induced signal contamination remains a major technical barrier to advance its clinical application.

\noindent
\textbf{New method:} Existing artifact-removal strategies, both front-end and back-end, face trade-offs between artifact suppression and algorithmic flexibility. Among back-end algorithms, Shrinkage and Manifold-based Artifact Removal using Template Adaptation (SMARTA) has shown promising performance in mitigating stimulus artifacts with minimal distortion to local field potentials (LFPs), but its high computational demand and inability to handle transient direct current (DC) artifacts limits its use in real-time applications. To address this, we developed SMARTA+, a computationally efficient extension of SMARTA capable of suppressing both stimulus artifacts and DC transient artifacts while supporting flexible algorithmic design.

\noindent
\textbf{Results:} We evaluated SMARTA+ using semi-real aDBS data and real data from Parkinson’s disease patients. It preserved the spectral and temporal structure of the underlying LFPs and demonstrated robustness across a variety of simulated stimulation protocols. Furthermore, the temporal event localization analysis showed that SMARTA+ can accurately determine the beta-burst events.

\noindent
\textbf{Comparison with existing methods:}
SMARTA+ outperformed template subtraction, pulse blanking, and transient blanking, and achieved performance comparable to or better than SMARTA while substantially reducing computation time.

\noindent
\textbf{Conclusions:} By enhancing artifact suppression and improving computational efficiency, we show that SMARTA+ has the potential to advance real-time, closed-loop aDBS systems for neuromodulation therapies across diverse neurological disorders.
\end{abstract}
\maketitle
\markboth{\shorttitle}{\shorttitle}

\section{Introduction}
Deep brain stimulation (DBS) is a widely used and effective therapy for advanced Parkinson's disease (PD) \cite{limousin1995DBS,benabid2003DBS,krack2003DBS,deuschl2006DBS}. By delivering electrical stimulation through implanted electrodes in the subthalamic nucleus (STN), DBS can alleviate motor impairments and improve patients' quality of life. Additionally, local field potentials (LFPs) can be recorded from the same electrodes during stimulation, providing real-time insights into PD-related neural activity. Biomarkers derived from LFP enable adaptive deep brain stimulation (aDBS), where stimulation parameters are dynamically adjusted to respond to fluctuations in disease state \cite{little2013aDBS,arlotti2018aDBS8hour}. The clinical benefits of aDBS have been demonstrated in PD, with improvements observed in motor symptoms such as tremor \cite{velisar2019aDBS} and bradykinesia \cite{little2016aDBS}. Beyond its application in PD, aDBS has also shown promise in treating essential tremor \cite{yamamoto2013ET, opri2020ET} and epilepsy \cite{kossoff2004aDBSinEpilepsy,stypulkowski2014aDBSinEpilepsy}. These studies suggest that aDBS may be broadly applicable to various movement disorders, offering a more efficient and patient-specific therapeutic strategy.

Despite its promise, the clinical implementation of adaptive deep brain stimulation (aDBS) remains limited by significant technical challenges, particularly contamination from stimulation-induced artifacts. Notably, the amplitude of artifacts generated by clinically effective stimulation (up to 3 V) can be several orders of magnitude greater than that of local field potential (LFP) signals, which typically reside in the microvolt ($\mu$V) range. This large amplitude disparity introduces complications such as aliasing, system nonlinearity, and even amplifier saturation, all of which hinder the reliable extraction of symptom-related biomarkers
 \cite{stanslaski2012aDBS}. 
Various techniques, spanning both front-end and back-end processing, have been developed to mitigate stimulation artifacts in aDBS recordings.

When the frequency band of interest is known in advance, careful selection of stimulation frequency and sampling rate (known as {\em frequency planning} or {\em interleaving}) \cite{thenaisie2021percept} is often considered to mitigate aliasing, especially in preserving the beta band for Parkinson's disease (PD) treatment \cite{little2013aDBS,arlotti2018aDBS8hour}. However, this approach has limitations, since the dynamic morphology of stimulation artifacts over time can still degrade signal quality via spectral leakage and aliasing, particularly in the presence of large artifacts, and thereby limit the potential for optimizing therapeutic efficacy. 
Moreover, even if beta band activity (13-35 Hz) can be well preserved, extending protection to other relevant frequency bands remains challenging, which constrains the flexibility of stimulation parameters and its generalization to other applications. Gamma (60-90 Hz) \cite{swann2018aDBSgamma,lofredi2018aDBSgamma} and alpha band activity (4-8 Hz) \cite{marceglia2011alpha} have both been proposed as potential biomarkers for aDBS in PD, and high-frequency oscillations (HFOs) above 200 Hz have been shown to be associated with tremors in PD \cite{hirschmann2016HFO}. Very high-frequency oscillation (VHFO) in the 1000-2500 Hz range \cite{usui2015VHFO} have shown utility in epilepsy and may also hold relevance for PD. 
Furthermore, accurately removing stimulus artifacts alone may not suffice. The abrupt switching of stimulation often induces direct current (DC) transient artifacts at stimulation onset and offset, which can further contaminate biomarkers \cite{anso2022artifactRemoval}. Together, these artifact-related challenges not only compromise the fidelity of LFP-based biomarkers but also impair aDBS control algorithms by introducing erroneous feedback and triggering unintended stimulation \cite{neumann2023aDBSreview}. These observations underscore the need for robust and flexible artifact removal algorithms capable of preserving a broad spectrum of physiologically meaningful signals.

In general, a basic front-end approach is common-mode rejection, where two sensing contacts adjacent to the stimulation electrode are used to obtain bipolar LFPs by signal subtraction \cite{little2013aDBS,arlotti2018aDBS8hour}. Ideally, when contacts are symmetrically positioned around the stimulation site, artifacts cancel out. However, impedance mismatches and tissue asymmetries often leave residual artifacts and DC transient distortions \cite{anso2022artifactRemoval}. To further reduce artifacts, blanking techniques can be applied at the front-end \cite{stanslaski2018aDBSdevice} or back-end \cite{heffer2008interpolation}, temporarily disabling signal acquisition during stimulation pulses. While effective at suppressing high-amplitude artifacts, blanking also removes underlying neural signals and fails to address transient DC artifacts that persist beyond the stimulation period. 
%
%When the frequency band of interest is known in advance, such as the beta band in PD therapy \citep{little2013aDBS,arlotti2018aDBS8hour}, careful frequency planning is a commonly used strategy with limtation, as previously discussed. %
Digital filtering and template subtraction (TS) are widely used back-end methods. TS assumes artifact stability over time, averaging repeated instances to subtract from the signal  \cite{hashimoto2002templateSub,sun2016MVS}. However, time-varying brain responses may alter artifact characteristics \cite{liu2024}, limiting TS performance.
%These challenges highlight the need for more robust artifact removal techniques to ensure reliable feedback signals for aDBS.
To address this, Shrinkage and Manifold-based Artifact Removal using Template Adaptation (SMARTA) was introduced \cite{liu2024}. By leveraging manifold denoising and random matrix theory, SMARTA adapts templates for individual artifacts using the underlying geometry of stimulation-induced distortions and the high-dimensional nature of LFP noise. This enables precise artifact removal while preserving neural signals, outperforming existing algorithms.

Despite its strengths, SMARTA has two key limitations. First, it is computationally intensive: processing a 7-ms segment under 22 kHz sampling rate takes over 200 ms, making it unsuitable for real-time aDBS. Second, like TS, it does not resolve DC transient artifacts, which limits its use for short or repeated stimulation periods and impairs beta burst detection. To support real-time aDBS with an accurate beta burst detection, a more efficient and DC-transient-aware variant of SMARTA is needed for accurate and timely biomarker extraction. Although the theoretical framework of SMARTA allows for signals with heterogeneous sampling rates and stimulation frequencies, a systematic evaluation of this capability remains lacking. Moreover, to our knowledge, a reliable detection of dynamic events, for instance, the onset and offset of beta bursts in PD patients, is less discussed in the literature.

\begin{figure}[!htb]
    \centering
    \includegraphics[width=0.9\textwidth]{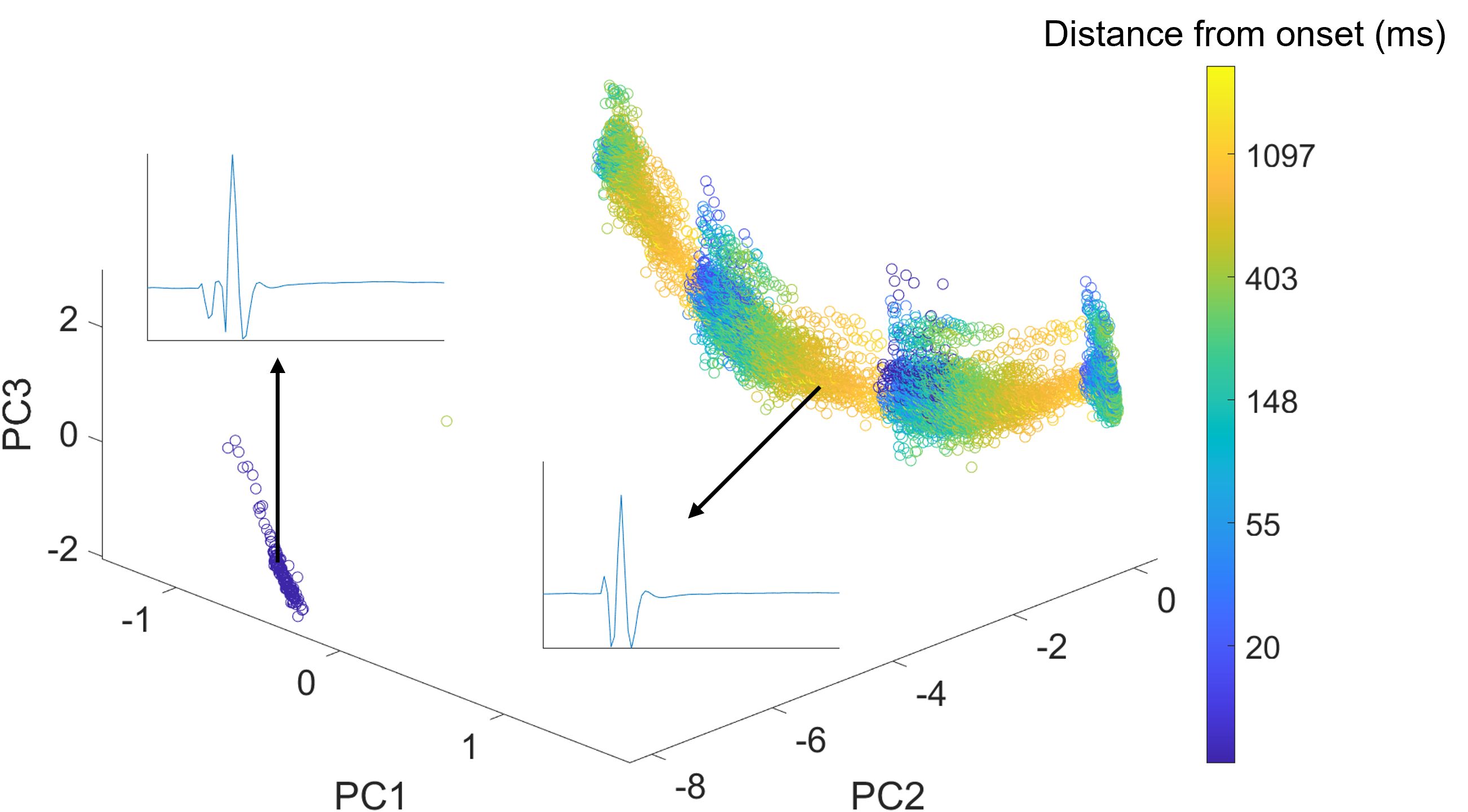}
    \caption{\label{fig:pca} 
Visualization of stimulus artifacts recorded during aDBS using principal component analysis (PCA), with the top three principal components shown. The color bar indicates the timing of each artifact relative to stimulation onset. Notably, artifact patterns near the onset differ markedly from those at later times, revealing changes in artifact dynamics over time. Since PCA is a linear dimensionality reduction method, the observed structure suggests underlying nonlinear structure and nontrivial dynamics among the artifacts.}
\end{figure}

In this study, we refined SMARTA to address its limitations, resulting in SMARTA+, which preserves SMARTA's core strengths while greatly improving computational efficiency, mitigating DC transient artifacts and reliably detecting beta bursts. A key enhancement involved replacing SMARTA's most computationally intensive step, the k-nearest neighbors (KNN) search, with an approximate nearest neighbors (ANN) algorithm \cite{ANNOY,jones2011RANN,jones2013RANN}. Unlike KNN, which performs exhaustive searches, ANN uses decision trees to identify similar artifacts with significantly reduced computational burden \cite{ANNOY}. Additionally, we introduced a dimensionality reduction step based on wavelet transforms, which exploits the sparsity structure of stimulus artifacts \cite{donoho1995wavelet}. These modifications enable the construction of a large stimulus artifact library, potentially from other subjects, allowing SMARTA+ to match each artifact with the most similar examples across datasets. This cross-subject learning improves artifact estimation and suppression, especially during stimulation onset, where DC transients alter artifact morphology (see Figure \ref{fig:pca}). Without a sufficiently diverse library, onset artifacts are particularly difficult to model. To further suppress DC transients, we incorporated a line-fitting algorithm that estimates and removes the slow-varying baseline shift at the start of each stimulation period. Together, these advances enable SMARTA+ to achieve effective artifact removal with a fraction of the original computational cost, advancing it toward real-time aDBS deployment.

We evaluated the performance of SMARTA+ using aDBS signals recorded from PD patients and semi-real aDBS signals synthesized from these recordings. 
For benchmarking, we compared its performance against template subtraction \cite{culaclii2016ts}, SMARTA \cite{liu2024} and blanking methods \cite{anso2022artifactRemoval}.
In addition to standard evaluation metrics, including the normalized mean square error (NMSE), artifact residual (AR), and spectral concentration (SC), which respectively quantify LFP recovery across spectral bands, in the time domain, and in the spectral domain \cite{liu2024}, we introduced a temporal event localization analysis to assess how accurately SMARTA+ recovers beta burst events. Specifically, we evaluated recall, precision, and F1-score to quantify the detection of beta burst onset and offset. This temporal event localization analysis, while previously applied in sEMG-based closed-loop DBS systems  \cite{shukla2012} for evaluating closed-loop DBS with post-stimulation control signals and sleep apnea detection from the photoplethysmogram signal \cite{chen2025validation}, have not yet been used to evaluate aDBS algorithms that rely on real-time beta activity from LFPs. To our knowledge, SMARTA+ is the first algorithm explicitly designed to preserve beta burst timing. 

\section{Materials} \label{bigSec:material}
% Patients undergoing bilateral implantation of STN-DBS for the treatment of advanced PD were recruited. The surgical procedures were described in \citep{chen2021YCC}. LFPs were recorded one day after electrode implantation through the externalized electrode using Neuro Omega (Alpha Omega Engineering, Israel). All surgeries and recordings were conducted at Chang Gung Memorial Hospital, Linkou, Taoyuan, Taiwan, between 2023 and 2024 by the coauthors Dr. P. H. Tu and Dr. M. C. Yeap.
Patients with advanced Parkinson’s disease who underwent bilateral STN-DBS implantation were included through secondary use of clinical data obtained from prior IRB-approved studies. The surgical procedures have been described previously \cite{chen2021YCC}. Local field potentials (LFPs) were recorded one day after electrode implantation using externalized leads and the Neuro Omega system (Alpha Omega Engineering, Israel). All surgeries and recordings were performed at Chang Gung Memorial Hospital, Linkou, Taoyuan, Taiwan, between 2023 and 2024 by the coauthors Dr. P. H. Tu and Dr. M. C. Yeap.

The experimental procedure for DBS data collection began with baseline LFP recording. The sampling rate was 22 kHz, and each recording lasted 200 seconds. The beta peak amplitudes in the bipolar LFP spectrum were derived from contact pairs 0-2 and 1-3 (model 3389, Medtronic). The contact pair exhibiting the highest beta peak was selected for stimulation. Next, the impedance of the stimulation contact was measured, and the maximum stimulation current was determined, ensuring that the equivalent voltage remained below 3.6 V. The stimulation frequency was set at 130 Hz, with a pulse width of 60 $\mu$s. 

For cDBS recordings, stimulation started 10 seconds after session onset and ended 10 seconds before completion. The aDBS protocol, particularly the threshold, is summarized in the next subsection.  
LFPs were recorded using the Neuro Omega system and processed in real time in MATLAB 2019a. Stimulation was triggered if beta amplitude exceeded the threshold and remained on for at least 400 ms; it was deactivated only when the amplitude fell below the threshold. For each patient, aDBS was recorded for 120 s using the same current amplitude as cDBS.

\subsection{aDBS protocol} \label{sec:beta extraction}
The aDBS protocol followed the method of \cite{little2013aDBS}.
A second-order Butterworth band-pass filter (3-37 Hz) was first applied to the baseline LFP signal. The peak beta frequency (13-35 Hz) was identified, and a peak IIR filter with a Q factor of 3 centered at this frequency was applied three times to enhance beta activity. The resulting signal was then rectified and smoothed using a 400-ms moving average to extract {\em beta amplitude}. Stimulation was driven by a threshold set at the 75th percentile of the amplitude distribution. During the aDBS procedure, the online beta amplitude was calculated using the same method. Stimulation was initiated when the amplitude exceeded the predefined threshold and was turned off once it fell below that threshold. A minimum stimulation duration of 400 ms was enforced.

\section{Proposed algorithm -- SMARTA+} \label{bigSec:method}

The SMARTA+ algorithm builds on the previously proposed SMARTA  (Shrinkage and Manifold-based Artifact Removal using Template Adaptation) framework \cite{liu2024}. SMARTA+ introduces two key enhancements. First, to reduce computational burden, the original KNN search is replaced with an ANN algorithm, and dimensionality is reduced via wavelet transform. These updates enable subject-adaptive, online construction of a larger and more diverse stimulus artifact library, improving artifact removal accuracy and enabling real-time implementation. Second, a curve-fitting procedure is incorporated to better handle transient DC artifacts.

\begin{figure}[!htb]
    \centering
    \includegraphics[width=0.8\textwidth]{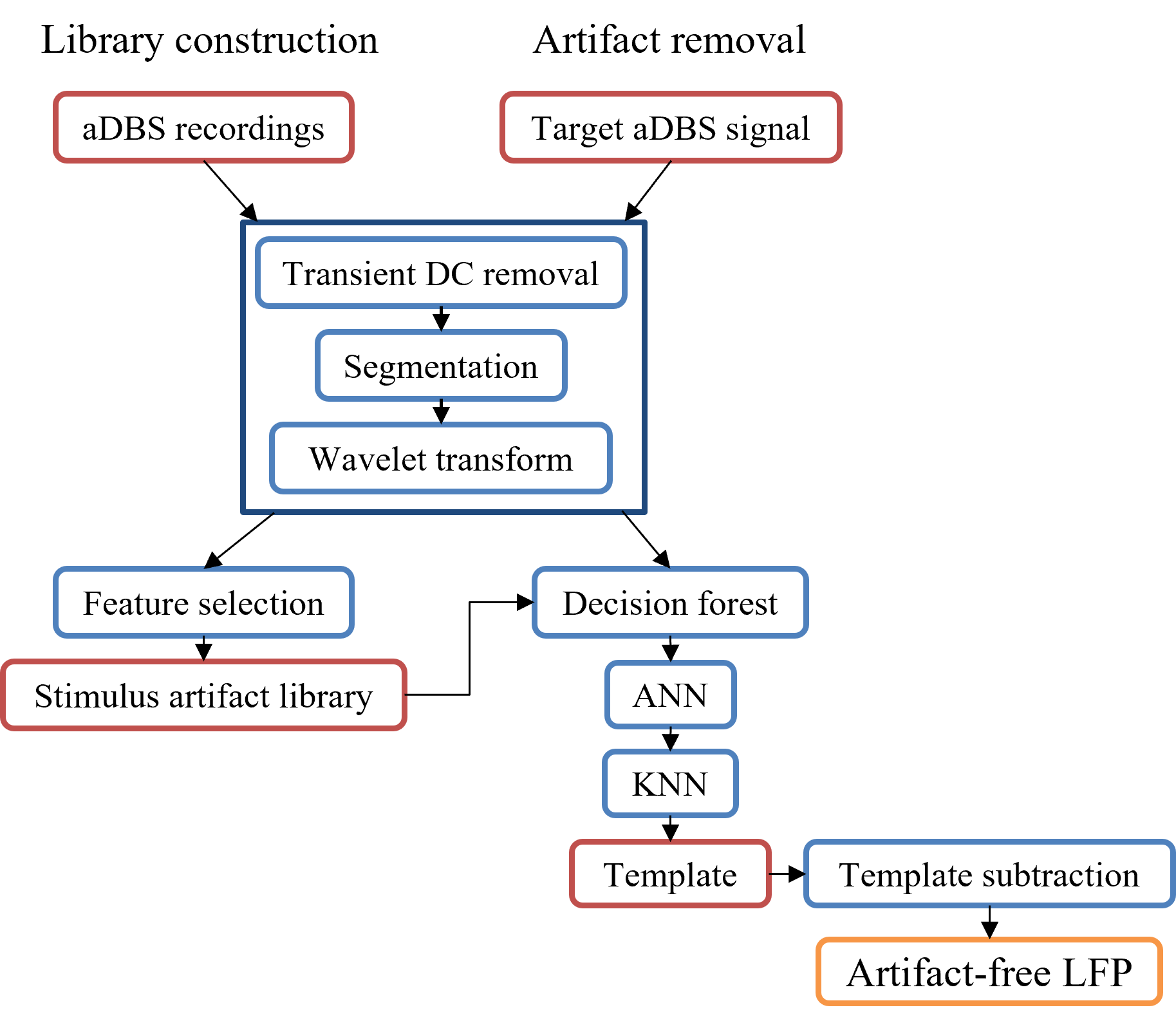}
    \caption{\label{fig:flowchart} Flowchart of SMARTA+.}
\end{figure}

The flowchart of SMARTA+ is shown in Fig.~\ref{fig:flowchart}. The overall workflow is as follows. A curve-fitting procedure is applied for all recordings to address transient DC artifacts (Section \ref{sec:DCremoval}). 
Prior to removing artifacts from a new recording, indexed by $n+1$, a {\em stimulus artifact library} is constructed from all available $n$ recordings (Section \ref{section library construction}), potentially including data from different subjects. With the library, each stimulus artifact in the $(n+1)$th recording is recovered by applying the following algorithm, which is a modification of SMARTA. %For each recording we want to eliminate the stimulus artifacts, 
First, construct a decision forest from the library, where the wavelet transform is applied. For each target stimulus artifact in the $(n+1)$th recording, similar artifacts are retrieved from the library using the constructed decision forest and ANN (Section \ref{sec:forest}), where the leaf nodes at the end of the decision paths form a candidate template pool. Within this pool, the $K$ nearest neighbors are identified, with the optimal $K$ selected via greedy search (Section \ref{sec:testing}). The final artifact template is computed as the median of these neighbors and subtracted from the signal segment. After removing stimulation artifacts, the same curve-fitting step is applied again to improve the residual transient DC artifact  removal (Section \ref{sec:DCremoval}), yielding the final artifact-free LFP. 

Below, we detail the algorithm step by step. 
Denote the stimulation frequency by $f_\mathrm{sti}$, which is typically 130 Hz in DBS therapy for PD. Assume there are $M$ aDBS recordings available and we aim to recover the LFP of the $(M+1)$th aDBS recording.

\subsection{Transient DC artifact removal} \label{sec:DCremoval}
For each recording, transient DC artifacts were removed prior to the segmentation of stimulus artifacts. Over each stimulation period, the trend containing the transient DC artifact is estimated by the following two-step procedure. The first step is to apply a moving-average filter with a window length of 10 ms to eliminate large outliers in the signal, and the second step is to fit a sixth-order polynomial function to the smoothed signal. For non-stimulation periods, the trend was estimated and subtracted using the same procedures, except that a first-order polynomial was used for fitting.

\subsection{Stimulus artifact library construction}\label{section library construction}
The library is constructed in two stages: artifact detection and segmentation, followed by ANN preparation, including data cleanup and dimension reduction. In real-time use, the target recording under analysis is excluded.

\subsubsection{Stimulus artifacts detection and segmentation} \label{sec:segmentation}
For the $m$-th recording, where $m=1,\ldots,M+1$, the peak-finding algorithm described in \cite{liu2024} was applied to obtain accurate timing of stimulus artifacts. Before identifying peaks, a third-order high-pass Butterworth filter with a cutoff frequency of 300 Hz was used to remove low-frequency noise. Additionally, a moving-average filter with a window length of 1 s was applied to estimate the trend, which was subsequently subtracted from the signal. The absolute value of the signal was then taken, followed by smoothing using a moving-average filter with a window length of 0.5 ms. 
A sample point was identified as a peak if it was greater than its neighboring points and exceeded a predefined threshold. This threshold was set as the 95th percentile of all sample points. Furthermore, the interval between detected peaks had to be greater than $1/f_\mathrm{sti}-\tau$, where $f_\mathrm{sti}$ is the stimulation frequency and $\tau$ is a bias-handling factor. In this study, $\tau$ was set to 0.5 ms. The detected peaks marked the locations of stimulus artifacts for segmentation. For aDBS signals, stimulation periods were further identified by detecting intervals between artifacts that were longer than the minimum period set in the aDBS algorithm.

Before segmentation, second-order IIR notch filters were applied to remove 60-Hz line noise and its harmonics up to 3000 Hz, with a Q factor of 200. Afterwards, segmentation was performed, with each segment containing a single artifact. The $i$-th segment started 1 ms before and ended at $1/f_\mathrm{sti}-0.5$ ms after the detected artifact location and is denoted as $x_i$. A data matrix was then constructed as
\begin{equation}
\XX_m=\left[x_{m,1}\ x_{m,2}\ \ldots \ x_{m,n_m}\right]\in \RX^{p\times n_m},
\end{equation}
where $p\in \mathbb{N}$ represents the length of each segment, and $n_m$ is the number of segments. $\XX_m$ is a noisy data matrix, which can be written as $\XX_m=\boldsymbol{S}_m+\boldsymbol{N}_m$, where $\boldsymbol{S}_m\in \RX^{p\times n_m}$ represents the stimulus artifacts that we want to recover from $\XX_m$, and $\boldsymbol{N}_m\in \mathbb{R}^{p\times n_m}$ contains LFP, which is viewed as noise.

\subsubsection{Prepare data for ANN} \label{sec:training}
Instead of processing all segments in $\XX_m$ at once, they were divided into smaller groups, each containing $2p$ segments, and the LFP was cleaned up by eOptShrink. This grouping strategy was implemented to speed up the eOptShrink algorithm. For the sake of self-completeness, eOptShrink algorithm can be found in \ref{sec:cOS}. After LFP removal by eOptShrink, each segment underwent wavelet transformation using the Haar wavelet and we obtained the wavelet coefficient matrix:
\begin{equation}
\boldsymbol{W}_m=\left[w_{m,1}\ w_{m,2}\ \ldots \ w_{m,n_m}\right]\in \RX^{q\times n_m},
\end{equation}
where $q\in \mathbb{N}$ is the number of wavelet coefficients.
The average stimulation amplitude, denoted as $a_m:=\frac{1}{n_m}\sum_{j=1}^{n_m}\max x_j$, and the median stimulus artifact, denoted as $s_m\in \mathbb{R}^q$, were computed as key characteristics of the $m$-th recording for subsequent analysis. Feature selection for applying ANN and KNN was then performed as follows. First, a matrix of all median stimulus artifacts $s_m$ was constructed: $S=[s_1\ s_2 \ldots s_M]\in \RX^{q\times M}$. 
For each $i=1,\ldots,q$, the variance of the $i$th row of $S$ was computed and denoted as $s_v(i)$, yielding a variance vector $s_v\in \mathbb{R}^q$.
The entries of $s_v$ were then sorted in descending order, and their cumulative sum was computed, resulting in a vector $\tilde{s}_v\in \mathbb{R}^q$. The index $k$ was defined as the largest integer such that $\tilde{s}_{v}(k)<0.99\tilde{s}_{v}(q)$. The indices of the top $k$ entries in the sorted array were then selected as features for the subsequent ANN and KNN analyses.

\subsection{Build decision forest} \label{sec:forest}
For the $(M+1)$th aDBS signal, recordings were selected from the other $M$ recordings with stimulation amplitude different from $a_{M+1}$ by less than 10\%; that is, all $m$ s.t. $\left|\frac{a_{m}-a_{M+1}}{a_{M+1}}\right|\leq 0.1$ were selected as candidates. Among these candidates, the KNN algorithm was applied to identify $Q$ recordings with closest $s_m$ to $s_{M+1}$. If fewer than $Q$ candidates were available, all candidates were included. Collect all stimulus artifacts as $\GX_{M+1}=\left[w_1\ w_2 \ \ldots\  w_{n'}\right]\in \mathbb{R}^{q\times n'}$, where $n'$ represents the number of segments in the selected $K$ recordings. 

Decision trees were then constructed using $\GX_{M+1}$. To build a tree, two points $w_i$, $w_j$ were randomly selected to generate a hyperplane, which is determined by the middle point $\frac{w_i+w_j}{2}$ with the normal direction $\frac{w_i-w_j}{\|w_i-w_j\|}$. The data were then divided into two subspaces by the hyperplane. Once a subspace contained fewer than $M'$ points, it was not further partitioned. Otherwise, this process was repeated until all subspaces contained fewer than $M'$ elements, thus yielding a tree structure. The resulting subspaces were termed {\em end nodes} of the tree, and the data points they contained were referred to as {\em leaves} of the tree. Since the selection of points for hyperplane derivation was random, the results from a single decision tree could be unstable and potentially inaccurate. To enhance accuracy, $T$ trees were constructed, forming the {\em initial decision forest}. While a larger $T$ improved accuracy and stability, it also increased computational cost.

During the first stimulation period, SMARTA+ employed the initial decision forest. Upon completion of each stimulation period, artifacts extracted from that period were added to the artifact library, after which the decision forest was reconstructed online according to the method outlined previously. This iterative update enabled the forest to gradually adapt to the target signal.

\subsection{Derive template and recover LFPs} \label{sec:testing}
For each artifact segment, the corresponding wavelet coefficients were obtained. These coefficients were then used to locate the corresponding leaves in the decision forest. Specifically, for each node in a decision tree, the subspace corresponding to the wavelet coefficient set was determined by the hyperplane of that node. The set continued traversing the tree until it reached an end node, where the contained leaves were considered its neighbors. After searching through the entire forest, the ANN algorithm identified at most $M\times N$ neighbors, collectively denoted as $\YX$.

Next, the KNN was applied to $\YX$, selecting $K$ closest neighbors. The final template for the $i$th stimulus artifact, denoted as $\hat{\boldsymbol s}^*_i$, was then computed by taking the entry-wise median of the $K$ chosen neighbors. The optimal value of $K$ was determined using a greedy search algorithm that minimized the artifact residual (AR) index \cite{malik2017ARindex}, defined as
\begin{equation} \label{eq:ar}
\begin{split}
\mathrm{AR}_i=&\left|\mathrm{log}\left(\frac{1}{2}\left[\frac{\mathrm{med}|\hat{Z}_i-\mathrm{med}(\hat{Z}_i)|}{\mathrm{med}|Z_i-\mathrm{med}(Z_i)|}+\frac{\mathrm{med}|Z_i-\mathrm{med}(Z_i)|}{\mathrm{med}|\hat{Z}_i-\mathrm{med}(\hat{Z}_i)|}\right]\right)\right.\\
    &\left.\times \frac{1}{2}\left[\frac{\mathrm{max}|\hat{Z}_i-\mathrm{med}(\hat{Z}_i)|}{\mathrm{max}|Z_i-\mathrm{med}(Z_i)|}+\frac{\mathrm{max}|Z_i-\mathrm{med}(Z_i)|}{\mathrm{max}|\hat{Z}_i-\mathrm{med}(\hat{Z}_i)|}\right]\right|,
\end{split}
\end{equation}
where $\hat{Z}_i$ represents the first $t_1$ ms and $Z_i$ the last $t_2$ ms of the estimated LFP signal $\boldsymbol{x}_i-\hat{\boldsymbol s}_i^*$ obtained after template subtraction. The functions $\mathrm{med}$ and $\mathrm{max}$ denote the median and the maximum, respectively. In this study, $t_1=3$ ms and $t_2=5.2$ ms. The optimal $K$ was selected from the candidate values $K=5,15,\ldots,55$ to minimize the AR index.

The overlap-and-add method was then applied to reconstruct the signal, using a window function defined as 
\begin{equation}
    w(t)=\left\{ \begin{array}{ccl}
         \sin^2\left(\frac{\pi(t-1)}{2g_1}\right) & \mathrm{for} & t\leq g_1  \\
         1 & \mathrm{for} & g_1 < t \leq g_2 \\
         \cos^2\left(\frac{\pi(p-t)}{2g_2}\right) & \mathrm{for} & g_2 < t \leq p
    \end{array} \right. ,
\end{equation}
where $g_1$ and $g_2$ represent the number of overlapping points with the preceding and following segments, respectively. The final artifact-free LFP signal was obtained by subtracting the sequence of artifact templates from the original signal. Additionally, signal trend removal was performed using the smoothing and polynomial-fitting method described in Sec.~\ref{sec:DCremoval}. Notably, stimulation periods could be identified prior to artifact removal.

\subsection{Performance evaluation} \label{sec:measures}

\subsubsection{Semi-real aDBS recording generation} \label{sec:semiReal generation}
To compare different stimulus artifact removal algorithms, semi-real aDBS signals were generated using the set of stimulus artifacts extracted in Sec.~\ref{sec:training} for each aDBS signal, and a decision forest for ANN was constructed as described in Sec.~\ref{sec:forest}. After artifact detection, trend removal (with the trend preserved for later use), and segmentation, the semi-real stimulus artifact for each segment was generated following the procedure in Sec.~\ref{sec:testing}, with the modification that the number of neighbors $K$ for KNN was fixed at 500. Additionally, instead of taking the median, the final result was computed by averaging all neighbors. 

With a large $K$, not all selected neighbors closely resembled the target artifacts, but this approach effectively removed the underlying LFP. The overlap-and-add method was then applied to reconstruct the semi-real artifacts, and the previously extracted trends for both stimulation and non-stimulation periods were reintroduced. As a result, the generated signal retained the stimulus artifacts and the trend of the original aDBS signal while minimizing the presence of underlying LFP activity. Finally, the semi-real aDBS signal was obtained by combining the artifact-only signal with an LFP signal recorded in the absence of stimulation.

In clinical adaptive DBS systems, the neural sensing sampling rate is typically around 1 kHz \cite{cummins2021DBSfs}. Accordingly, the semi-real recordings were resampled to 1.1 kHz to evaluate SMARTA+ under a clinic-like acquisition setting.

To further assess SMARTA+'s capability of handling different clinical stimulation parameters, the stimulation frequency was adjusted. Motivated by stimlation parameters used in DBS \cite{cagnan2019DBS180}, after constructing the semi-real artifacts as described above, artifact-only signals at a 180 Hz stimulation rate were added.

\subsubsection{Evaluation measures}
The effectiveness of artifact removal was quantified using three indices: artifact residual (AR), spectral concentration (SC), and normalized mean square error (NMSE), as described in \cite{liu2024}. The AR for the entire signal was computed by averaging the AR index, $\mathrm{AR}_i$, across all segments ($i=1,...,N$, where $N$ was the total number of segments). The AR index for each segment, $\mathrm{AR}_i$, was defined in Eq.~\ref{eq:ar}, where $Z_i$ was formed by concatenating the last $t_2$ ms of the $(i-20)$th segment to the $(i+20)$th segment. 

The SC index, which evaluated signal concentration at a target frequency $f_c$,  was calculated as
\begin{equation}
    \mathrm{SC}(f_c)=\frac{\sum_{f\in F_1}P_{\hat{Z}}(f)}{\sum_{f\in F_2}P_{\hat{Z}}(f)},
\end{equation}
where $P_{\hat{Z}}$ represents the power spectral density of the estimated signal $\hat{Z}$. The two frequency sets were defined as $F_1=\{f|f_c-1\leq f\leq f_c+1\}$ and $F_2=\{f|f_c-20\leq f\leq f_c+20\}$. 

In the semi-real setup, the NMSE was calculated as
\begin{equation}    \mathrm{NMSE}=10\mathrm{log}_{10}\frac{\sum_n (\hat{Z}(n)-Z(n))^2}{\sum_n Z(n)^2},
\end{equation}
where $\hat{Z}$ and $Z$ were the estimated and the ground-truth LFP signals, respectively. 

\begin{figure}[!htb]
    \centering
    \includegraphics[width=0.9\textwidth]{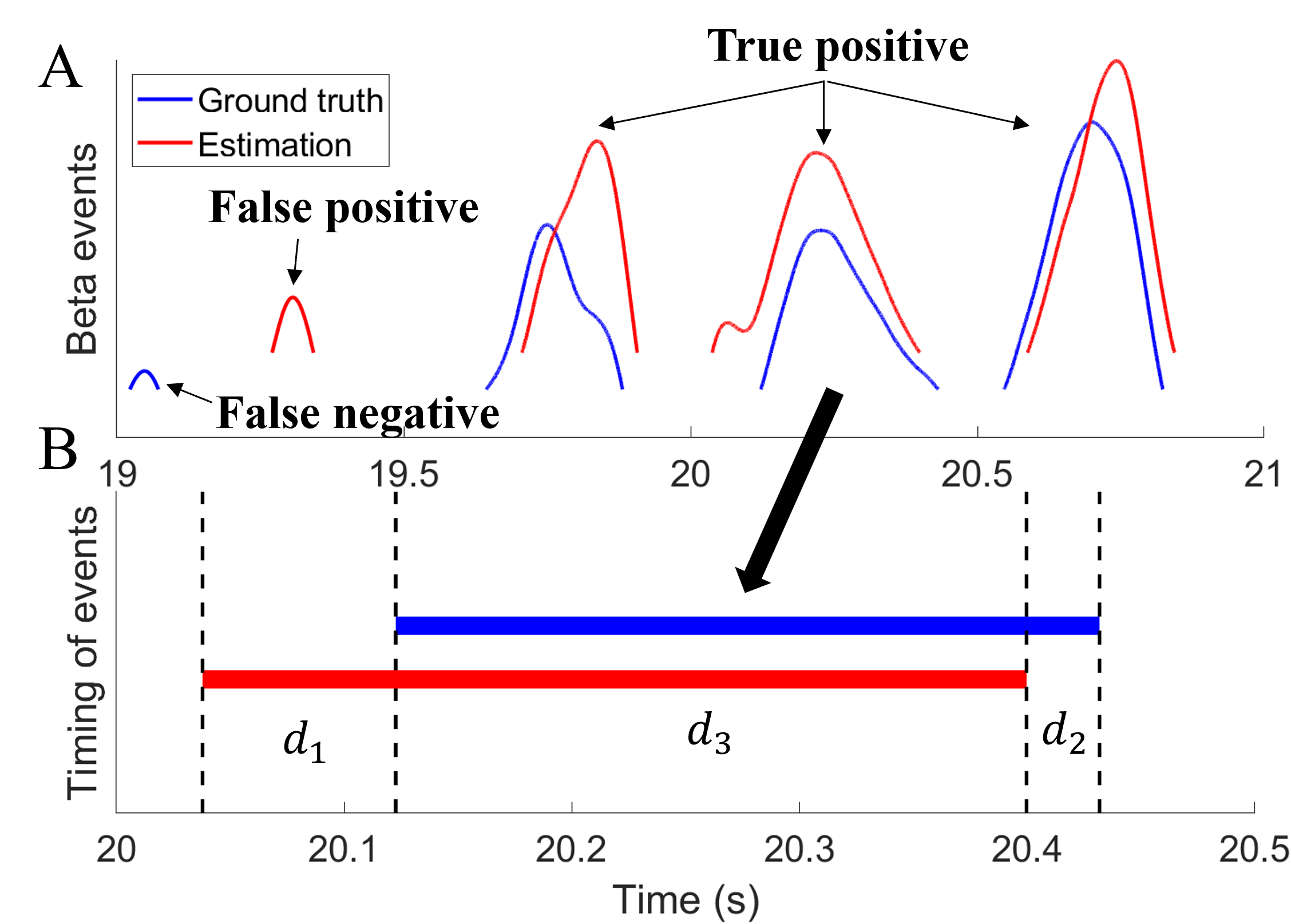}
    \caption{\label{fig:TPTNFN}
    Performance evaluation for the aDBS algorithm. (A) Illustration of true positive, false negative, and false positive beta events. (B) Definition of deviation $d_1+d_2$ with the unit s, and overlap ratio OR$_1=d_3/(d_1+d_3)$ and OR$_2=d_3/(d_2+d_3)$} with no unit, of a true positive event.
\end{figure}

In addition to these indices, the accuracy of beta burst detection, or temporal event localization analysis, using the aDBS algorithm combined with SMARTA+ was evaluated using semi-real aDBS signals. The aDBS algorithm is described in Sec.~\ref{sec:beta extraction}. The classification of true positive (TP), false negative (FN), and false positive (FP) beta bursts is illustrated in Figure \ref{fig:TPTNFN}A. A beta burst detected by the algorithm was referred to as a beta event. A detected event from the estimated LFP was defined as TP if it overlapped with a ground truth beta event, and its deviation was measured as the total time difference between the estimated and ground truth events. As illustrated in Figure \ref{fig:TPTNFN}B, the deviation for a TP event was calculated as $d_1+d_2$ and two overlap ratios were calculated, including OR$_1=d_3/(d_1+d_3)$ and OR$_2=d_3/(d_2+d_3)$. A ground truth beta event is defined as FN when it does not overlap with any detected events in the estimated LFP, while a detected beta event from the estimated LFP is defined as FP if it does not overlap with any ground truth beta events. The total number of TP, FN, and  FP events were denoted as \#TP, \#FN, and \#FP, respectively.

For an aDBS algorithm, two key concerns were: (1) how accurately beta events were detected and (2) how many detected events were incorrect. To carry out the temporal event localization analysis, consider the following metrics. The first one was evaluated using recall, defined as 
$\mathrm{Recall}=\frac{\#\mathrm{TP}}{\#\mathrm{TP}+\#\mathrm{FN}}$.
The second concern was assessed using precision, defined as 
$\mathrm{Precision}=\frac{\#\mathrm{TP}}{\#\mathrm{TP}+\#\mathrm{FP}}$.
These two indices were further summarized using the F1-score, given by
$\mathrm{F1}=\frac{2\times\mathrm{Recall}\times\mathrm{Precision}}{\mathrm{Recall}+\mathrm{Precision}}$.
Recall quantified the proportion of true events detected, precision measured the proportion of detected events that are true, and the F1-score provided a balanced summary of recall and precision. 

\subsubsection{Other methods for comparison}
To validate the performance of SMARTA+, comparative analyses were conducted against SMARTA and two artifact removal techniques: template subtraction and blanking. SMARTA was implemented following the methodology described in \cite{liu2024}, without any modifications. Template subtraction was implemented based on the causal setup described in \cite{culaclii2016ts}, where templates are generated in real time using only past data. Specifically, for the $i$th segment, the template was computed as the average of the $(i-k)$th to $(i-1)$th segments, ensuring that the estimation relied solely on preceding information. Notably, although \cite{culaclii2016ts} employed a hybrid approach combining both front-end and back-end modules, only the back-end component of template subtraction could be implemented in this study due to data constraints.

Blanking was also implemented in two forms: pulse blanking and transient blanking. Front-end blanking, as commonly applied in embedded aDBS devices \cite{stanslaski2018aDBSdevice}, was realized using the sample-and-interpolate technique \cite{heffer2008interpolation} to perform pulse blanking. Each stimulation artifact pulse was blanked over a 2.5 ms window, with the removed samples interpolated to maintain continuity. Transient blanking, implemented according to the method proposed in \cite{anso2022artifactRemoval}, was applied at the level of entire stimulation periods rather than individual artifact pulses. Specifically, a 550 ms window following the onset of each stimulation period was blanked to suppress the broadband noise associated with DC transient artifacts. 

These four methods, including SMARTA+, SMARTA, template subtraction, and blanking, were evaluated under consistent conditions to assess their effectiveness in suppressing stimulus artifacts and preserving relevant neural signals in aDBS recordings.

\section{Results}

The SMARTA+ algorithm was implemented in MATLAB 2020a and executed on a machine with an AMD Ryzen 5 3500X 6-core processor, 16 GB RAM, and Windows 10.
In this study we have 34 recordings. For SMARTA+, the parameters were set to $Q=3$, $M'=500$ and $T=50$. To ensure general applicability, no parameter tuning was performed. Benchmark algorithms were implemented with their recommended parameter settings.

\subsection{Visualization of different algorithms}
\begin{figure}[!htb]
    \centering
    \includegraphics[width=\textwidth]{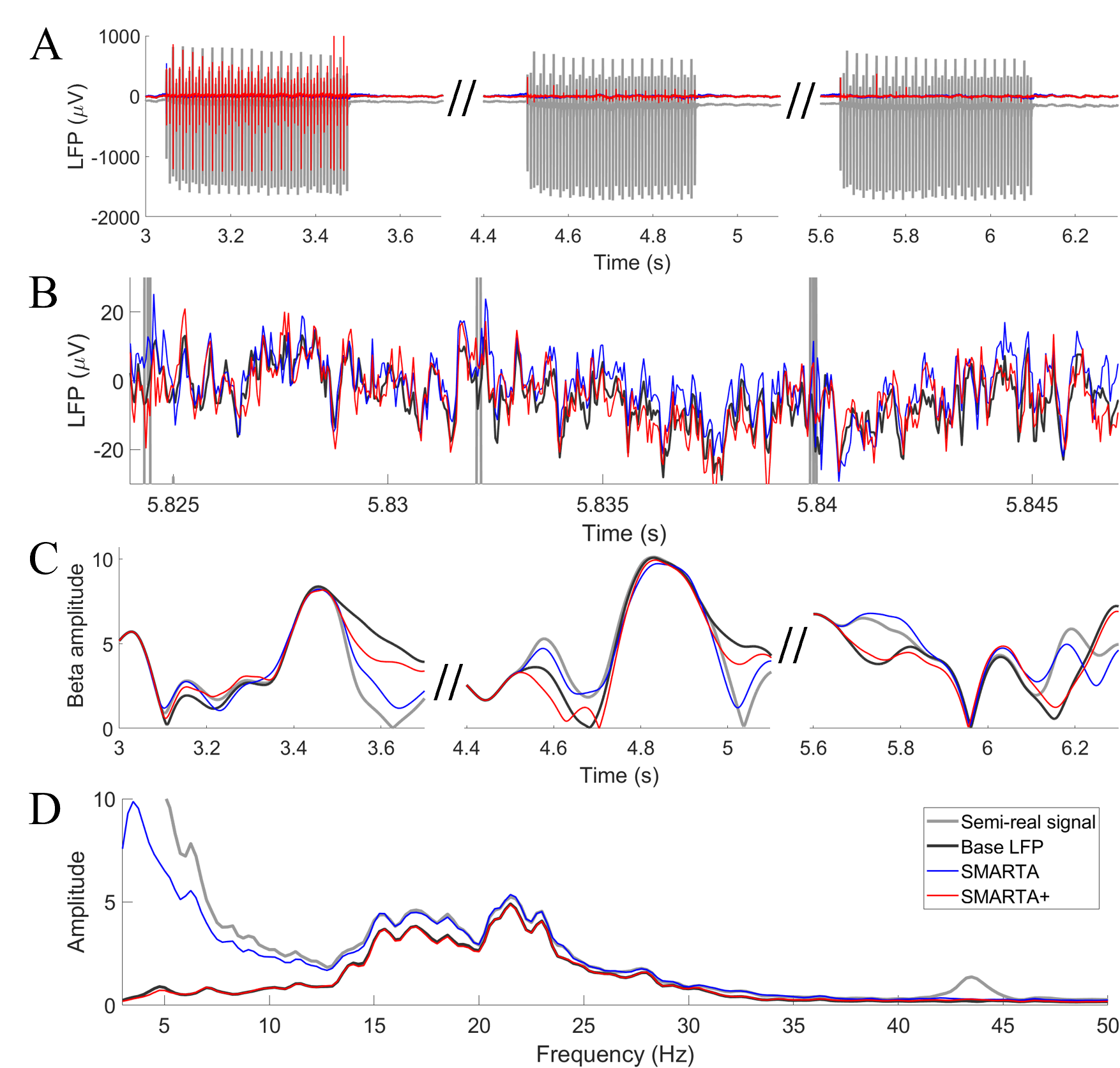}
    \caption{\label{fig:semiReal}
    Validation of SMARTA+ on a semi-real aDBS signal. Periods without stimulation are omitted to improve visualization. (A) Waveform of the semi-real aDBS signal (gray line), base LFP (black line), and artifact-free LFPs obtained using SMARTA (blue line) and SMARTA+ (red line). (B) Zoomed-in view of artifacts at the third stimulation period. (C) Beta amplitude estimated using the aDBS algorithm. (D) Power spectrum derived using Welch's method.}
\end{figure}

Building on prior reports that SMARTA outperforms template subtraction and blanking \cite{liu2024}, SMARTA+ was examined in both time and frequency domains for enhanced visualization and compared with SMARTA in this subsection. As shown in Figure \ref{fig:semiReal}D, trend-induced spectral spreading contaminated the beta band. Moreover, transient DC artifacts introduced beta-like activity at stimulation onsets and offsets (Figure \ref{fig:semiReal}C), distorting the beta amplitude, which was derived using the aDBS protocol described in Sec.~\ref{sec:beta extraction}, in the semi-real signal.  

In the time domain, SMARTA+ attenuated but did not fully eliminate stimulation artifacts in the first stimulation period because the initial library and decision forest came from out-of-signal exemplars (from other recordings), resulting in a template discrepancy. After each stimulation period, artifacts from the just-processed period were appended to the library and the decision forest was retrained, enabling template estimation from in-signal artifacts. This online adaptation made SMARTA+ progressively more stable and effective. As shown in Figure \ref{fig:semiReal}A, artifacts persisted in the first period but were effectively removed in the subsequent stimulation periods. A closer examination of the third stimulation period, shown in Fig. \ref{fig:semiReal}B, indicates that both SMARTA and SMARTA+ reliably reconstructed the underlying LFP signals, demonstrating stable performance after the initial adaptation phase, which typically requires only a few stimulation periods (approximately three).

In the frequency domain (Figure \ref{fig:semiReal}D), both SMARTA and SMARTA+ successfully removed the aliasing peak at 42.8 Hz, but SMARTA+ further reduced low-frequency noise compared to SMARTA since transient DC artifacts were removed and onset-related stimulus artifacts were better recovered. Regarding beta amplitude estimation, SMARTA failed to fully estimate contamination at stimulation onsets and offsets (Figure \ref{fig:semiReal}C). In contrast, SMARTA+ achieved better artifact removal at onsets and successfully removed trends after offsets, allowing the base LFP's beta amplitude to be recovered more accurately.

\begin{figure}[!htb]
    \centering
    \includegraphics[width=\textwidth]{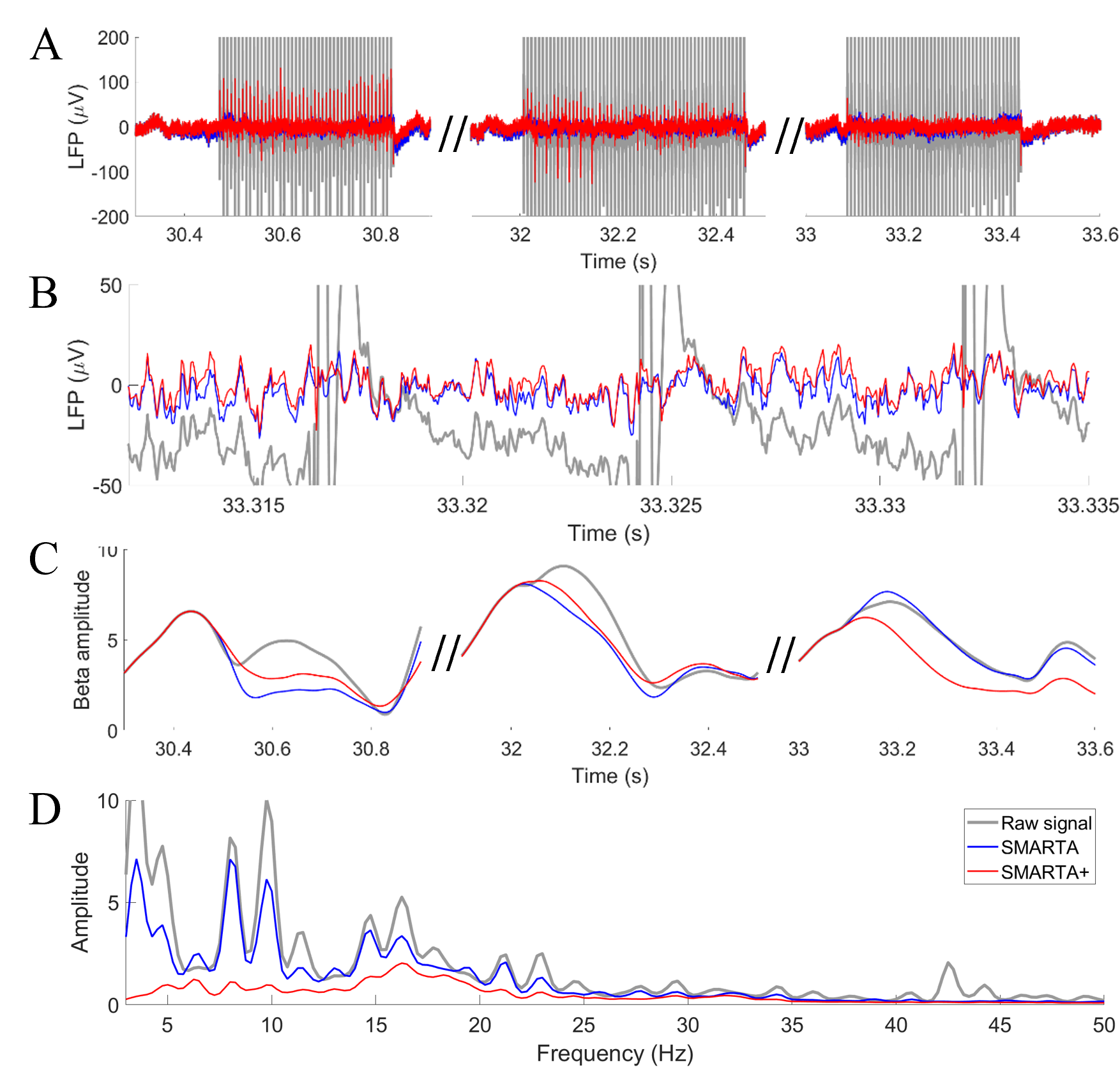}
    \caption{\label{fig:real}
   Validation of SMARTA+ on a real aDBS signal. Periods without stimulation are omitted to improve visualization. (A) Waveform of the real aDBS signal (gray line), and artifact-free LFPs obtained using SMARTA (blue line) and SMARTA+ (red line). (B) Zoomed-in view of artifacts at the third stimulation period. (C) Beta amplitude estimated using the aDBS algorithm. (D) Power spectrum derived using Welch's method. }
\end{figure}

Figure \ref{fig:real} presents a comparison using a real aDBS signal, demonstrating results consistent with those observed in the semi-real case. SMARTA+ was unable to fully remove artifacts during the first two stimulation periods, after which its performance stabilized and remained effective. In the frequency domain, SMARTA+ achieved greater suppression of low-frequency noise. Beta amplitudes were significantly distorted at the stimulation onsets and offsets; SMARTA partially reduced these effects, whereas SMARTA+ substantially attenuated them.

\subsection{Performance evaluation}

SMARTA+, SMARTA, template subtraction, and blanking were evaluated using 34 real and semi-real aDBS recordings. A quantitative comparison of these methods, alongside template subtraction and blanking, is presented in Table~\ref{tab:semiReal} and Table~\ref{tab:real} for semi-real signals and real signals, respectively. The accuracy of the estimated artifact-free signals was assessed using NMSE, and artifact removal effectiveness was evaluated using AR in the time domain and SC in the frequency domain. The AR values presented in Table~\ref{tab:semiReal} and Table~\ref{tab:real} represent the median across all signal segments, while SC values are averaged over specific frequency points: $f_c=129.16\times n$ Hz, $42.8+129.16\times m$ Hz, and $86.36+129.16\times m$ Hz for $n=1,2,3$ and $m=0,1,2,3$. The table also lists the average computation time per segment, denoted as $T_c$. 
Recall that due to the construction, SMARTA+ achieved stable performance after a few stimulation periods (See Figure~\ref{fig:semiReal}). To ensure a fair evaluation of the model's effective performance, in addition to results over the entire recording, we also report performance metrics with the first three stimulation periods excluded. The corresponding results are presented in the last row of the tables.

For semi-real signals, as shown in Table~\ref{tab:semiReal}, both SMARTA and SMARTA+ substantially outperformed the template subtraction and blanking methods across all metrics, except for SC values of SMARTA+ when the first three stimulation periods were included. Between the two, SMARTA+ yielded lower NMSE and AR value but slightly inferior SC value relative to SMARTA, suggesting comparable capability in restoring the underlying LFP signals. For real aDBS signals, as shown in Table~\ref{tab:real}, consistent trends were observed.
A significant advantage of SMARTA+ was its substantially lower computation time, decreasing the average processing time per segment by more than 90\% compared to SMARTA. This improvement enhances its feasibility for real-time aDBS applications.
The average time required to rebuild the decision trees in SMARTA+ was $346.24\pm 58.41$ ms, in addition to the artifact-removal computation time. The mean non-stimulation interval between successive stimulation periods was $588.08\pm 305.75$ ms, which allowed the decision-forest rebuilding procedure to be executed after each stimulation period and completed before the next.

\begin{table}[!htb] 
\centering
\begin{tabular}{ccccc} 
    \hline
    Median & \multirow{2}{*}{NMSE} & \multirow{2}{*}{AR} & \multirow{2}{*}{SC} & \multirow{2}{*}{$\mathrm{T}_c$ (ms)} \\
    MAD & & & & \\
    \hline
    \multirow{2}{*}{Raw data} & 25.0115 & 2.5192 & 0.4376 & N/A \\
    & (4.0334) & (0.2237) & (0.0938) & N/A \\
    \hline
    Template & 19.2288 & 1.5660 & 0.3791 & \textbf{0.0476} \\
    subtraction & (5.1384) & (0.3218) & (0.1106) & (0.0012) \\
    \hline
    Pulse & 12.5190 & 0.1372 & 0.4446 & 0.0947 \\
    blanking & (2.9617) & (0.1077) & (0.0558) & (0.0014) \\
    \hline
    Transient & 14.0001 & 0.4670 & 0.1369 & 0.1015 \\
    blanking & (5.3030) & ($<10^{-4}$) & (0.0810) & (0.0029) \\
    \hline
    \multirow{2}{*}{SMARTA} & -0.6043 & 0.0393 & \textbf{0.0577} & 83.7437 \\
    & (3.4507) & (0.0033) & (0.0044) & (1.1043) \\
    \hline
    SMARTA+ & -1.6195 & 0.0343 & 0.1814 & 2.2379 \\
    (whole signal) & (3.3831) & (0.0033) & (0.0565) & (0.0872) \\
    \hline
    SMARTA+ & \textbf{-9.0256} & \textbf{0.0333} & 0.0633 & 2.2439 \\
    (stable periods) & (7.7040) & (0.0030) & (0.0072) & (0.0816) \\
    \hline
    \end{tabular}
\caption{\label{tab:semiReal} Performance comparison of artifact removal methods on semi-real aDBS signals. MAD: median absolute deviation; NMSE: normalized mean square error; $\mathrm{NMSE}_\beta$: NMSE of $\beta$ band (13-35Hz); AR: artifact residual; $SC$: spectral concentration; $\mathrm{T}_c$: computation time per segment; N/A: not applicable; stable periods: excluding the first three stimulation periods for evaluation.}
\end{table}

\begin{table}[!htb] 
\centering
\begin{tabular}{cccc} 
    \hline
    Median & \multirow{2}{*}{AR} & \multirow{2}{*}{SC} & \multirow{2}{*}{$\mathrm{T}_c$ (ms)} \\
    (MAD) & & & \\
    \hline
    \multirow{2}{*}{Raw data} & 2.5175 & 0.4346 & N/A \\
    & (0.2278) & (0.0865) & N/A \\
    \hline
    Template & 1.5445 & 0.3702 & \textbf{0.0453} \\
    subtraction & (0.2855) & (0.1222) & (0.0008) \\
    \hline
    Pulse & 0.1375 & 0.4359 & 0.0905 \\
    blanking & (0.1049) & (0.0554) & (0.0005) \\
    \hline
    Transient & 0.4670 & 0.1354 & 0.0958 \\
    blanking & ($<10^{-4}$) & (0.0791) & (0.0008) \\
    \hline
    \multirow{2}{*}{SMARTA} & 0.0625 & \textbf{0.0894} & 83.0630 \\
    & (0.0135) & (0.0165) & (0.8342) \\
    \hline
    SMARTA+ & 0.0530 & 0.1892 & 3.5924 \\
    (whole signal) & (0.0131) & (0.0516) & (0.1127) \\
    \hline
    SMARTA+ & \textbf{0.0515} & 0.1073 & 3.5951 \\
    (stable periods) & (0.0129) & (0.0225) & (0.0961) \\
    \hline
    \end{tabular}
\caption{\label{tab:real} Performance comparison of artifact removal methods on aDBS signals.}
\end{table}

\subsection{Different spectral bands}
To further examine artifact removal performance across frequency bands, NMSE values were computed for alpha (4-8 Hz), beta (13-35 Hz), gamma (60-90 Hz), high-frequency oscillation (HFO; 200-400 Hz), and very high-frequency oscillation (VHFO; 400-1000 Hz and 1000-3000 Hz for VHFO$_1$ and VHFO$_2$, respectively) bands. The results are summarized in Table~\ref{tab:NMSE}. These values were obtained by applying band-pass filters to the signals before NMSE calculation. 
In the semi-real aDBS signals, the artifact signal was generated separately by averaging over 500 segments before added to the base LFP signal. This extensive averaging effectively removed most of the underlying beta-band activity from the artifacts, resulting in minimal distortion of beta activity in the raw semi-real data, as reflected by the low beta-band NMSE.
The alpha band exhibited high NMSE values in raw signals due to low-frequency drift caused by DC transient artifacts. Only SMARTA+, which incorporates a line-fitting algorithm for transient correction, demonstrated a significant reduction in NMSE in this band. In the gamma band, the aliasing of stimulation harmonics, particularly at 86.36 Hz, contributed to elevated NMSE in raw data. SMARTA+ again yielded the lowest NMSE, outperforming other compared algorithms. The HFO and VHFO bands contained a broader spectrum of artifact-induced distortion, including both harmonics and aliasing. As expected, the NMSE values for the raw semi-real data in the HFO and VHFO bands were the highest.
While transient blanking substantially improved NMSE in the HFO and VHFO bands compared with pulse blanking and template subtraction, SMARTA achieved the greatest overall reduction among these methods. When considering the stable periods of SMARTA+, its improvement in the HFO band exceeded that of SMARTA, while its performance in the VHFO band remained comparable. These findings confirm the superior performance of SMARTA+ in suppressing stimulation artifacts across time and frequency domains, yielding the most accurate restoration of LFP signals over the entire spectral range.

\begin{table}[!ht]
\footnotesize
\centering
\begin{tabular}{ccccccc} 
    \hline
    Med & \multirow{2}{*}{$\alpha$} & \multirow{2}{*}{$\beta$} & \multirow{2}{*}{$\gamma$} & \multirow{2}{*}{HFO} & \multirow{2}{*}{VHFO$_1$} & \multirow{2}{*}{VHFO$_2$}\\
    (MAD) & & & & & & \\
    \hline
    \multirow{2}{*}{Raw data} & 9.2380 & -3.4122 & 7.1040 & 21.9413 & 21.6738 & 26.6248 \\
    & (3.4241) & (4.4859) & (3.7075) & (3.8246) & (2.5157) & (3.5377) \\
    \hline
    Template & 9.0989 & -3.9024 & 5.0885 & 5.7281 & 8.5194 & 17.7320 \\
    subtraction & (3.3395) & (4.7417) & (4.3981) & (4.1816) & (4.0216) & (4.8221) \\
    \hline
    Pulse & 10.2108 & -1.8297 & 7.8926 & 18.1604 & 11.4267 & 15.3171 \\
    blanking & (3.9231) & (4.6398) & (3.7506) & (4.7542) & (4.0598) & (3.6732) \\
    \hline
    Transient & 13.7018 & 1.8069 & 5.0500 & 5.1466 & 1.7079 & 5.8455 \\
    blanking & (4.8620) & (3.3759) & (4.2632) & (6.2592) & (3.2128) & (3.6366) \\
    \hline
    \multirow{2}{*}{SMARTA} & 8.5588 & -6.0408 & 0.3658 & -6.9736 & -9.7279 & \textbf{-10.5621} \\
    & (3.4867) & (3.1311) & (2.8903) & (2.1999) & (2.8061) & (2.7886) \\
    \hline
    SMARTA+ & -5.4408 & \textbf{-13.5463} & -5.3779 & -1.9351 & -3.0674 & -2.0688 \\
    (whole signal) & (1.4129) & (3.1985) & (2.8050) & (6.0309) & (4.9093) & (4.0505) \\
    \hline
    SMARTA+ & \textbf{-5.8774} & -13.5198 & \textbf{-5.8394} & \textbf{-9.1458} & \textbf{-10.2569} & -10.3131 \\
    (stable periods) & (1.3845) & (3.2608) & (2.8709) & (3.6409) & (2.0767) & (1.4184) \\
    \hline
\end{tabular}
\caption{\label{tab:NMSE} Band-specific NMSE values for artifact removal methods across alpha (4-8 Hz), beta (13-35 Hz), gamma (60-90 Hz), HFO (200-400 Hz), VHFO$_1$ (400-1000 Hz), and VHFO$_2$ (1000-3000 Hz).}
\end{table}

\subsection{Temporal Event Localization Analysis}
Table~\ref{tab:TPFP} presents the accuracy of the beta event detection algorithm, using the baseline beta amplitude extraction procedure described in Sec.~\ref{sec:beta extraction}, following the application of various artifact removal methods. Below, ``Raw filtering'' denotes the condition in which no artifact suppression techniques was applied before evaluating beta amplitude. Transient blanking \cite{anso2022artifactRemoval} was intended to avoid artifact-induced self-triggering by suppressing beta detection for a fixed duration following stimulation onset. However, because it simply omitted the first 550 ms of each stimulation period, it did not contribute to accurate beta event detection and could not be meaningfully assessed. As such, transient blanking was excluded from direct performance comparisons. 

The performance metrics were averaged across 34 semi-real aDBS signals. To further investigate the efficacy of artifact removal at the initial phase of stimulation, detection performance was also evaluated separately within the first 50 ms of each stimulation period. This allows for a focused comparison of SMARTA+'s effectiveness at mitigating onset-related artifacts.

There are several findings. First, when no artifact removal was applied; that is, raw-filtering, the extracted beta amplitude resulted in high recall but low precision. This pattern was especially pronounced during the onset of stimulation, suggesting that many of the detected beta events were actually false positives caused by artifacts overlapping with the beta band. The inflated recall, therefore, reflected the detection algorithm's susceptibility to spurious high-amplitude signals, rather than its ability to identify true neural activity. 
Second, template subtraction, pulse blanking, and SMARTA all demonstrated improvements upon raw-filtering in detection accuracy by reducing the number of false positives. These method led to a notable increase in precision, thereby improving the F1-score and reducing timing deviation. Among them, SMARTA+ achieved the highest overall performance, particularly at the onset of stimulation. It demonstrated significantly higher precision and F1-score than other approaches, while maintaining a comparable recall. In addition, SMARTA+ notably reduced the timing deviation and increased the overlap ratio in beta event detection, providing more temporally precise alignment between the detected beta activity and the underlying neural signal.

The performance of SMARTA+ evaluated on the stable periods exceeded that obtained using the entire signal, suggesting that incomplete artifact suppression during the initial stimulation periods may introduce minor inaccuracies in beta-event detection. For most metrics, however, this difference was not statistically significant, and beta events were recovered with accuracy comparable to that observed during the stable phase.

Overall, SMARTA+ outperformed existing algorithms in several critical aspects. It preserved beta activity more effectively while maintaining strong artifact removal performance. Compared to SMARTA, its computational efficiency was significantly improved, making it more suitable for real-time applications. Furthermore, it enhanced the accuracy and timing of beta event detection in aDBS, particularly at stimulation onsets.

\begin{table}[!ht]
\scriptsize
\centering
\begin{tabular}{cccccccccccccc} 
    \hline
    & \multicolumn{2}{c}{Recall (\%)} & \multicolumn{2}{c}{Precision (\%)} & \multicolumn{2}{c}{F1-score (\%)} & \multicolumn{2}{c}{deviation (ms)} & \multicolumn{2}{c}{OR$_1$ (\%)} & \multicolumn{2}{c}{OR$_2$ (\%)}\\
     & Onset & All & Onset & All & Onset & All & Onset & All & Onset & All & Onset & All\\
    \hline
    \multirow{2}{*}{Raw filtering} & 91.33 & 91.70 & 57.30 & 77.25 & 68.27 & 83.54 & 172.37 & 99.80 & 71.89 & 81.30 & 87.95 & 90.21 \\
    & (0.19) & (0.02) & ($\ast$) & ($\ast$) & ($\ast$) & ($\ast$) & ($\ast$) & ($\ast$) & ($\ast$) & ($\ast$) & (0.02) & (0.001) \\
    \hline
    Template & \textbf{93.32} & 92.02 & 59.99 & 77.32 & 71.16 & 83.68 & 177.36 & 99.56 & 71.93 & 80.54 & 87.40 & 91.17 \\
    subtraction & (0.67) & (0.03) & ($\ast$) & ($\ast$) & ($\ast$) & ($\ast$) & ($\ast$) & ($\ast$) & ($\ast$) & ($\ast$) & (0.004) & (0.06) \\
    \hline
    \multirow{2}{*}{Blanking} & 93.17 & 91.40 & 63.23 & 78.16 & 73.45 & 83.98 & 175.60 & 101.76 & 72.48 & 80.88 & 86.64 & 90.33 \\
    & (0.80) & ($\ast$) & ($\ast$) & ($\ast$) & ($\ast$) & ($\ast$) & ($\ast$) & ($\ast$) & ($\ast$) & ($\ast$) & ($\ast$) & ($\ast$) \\
    \hline
    \multirow{2}{*}{SMARTA} & 92.73 & 91.83 & 63.10 & 78.82 & 73.21 & 84.50 & 163.59 & 95.34 & 74.09 & 81.66 & 88.27 & 90.48\\
    & (0.73) & (0.004) & ($\ast$) & ($\ast$) & ($\ast$) & ($\ast$) & ($\ast$) & ($\ast$) & ($\ast$) & ($\ast$) & (0.04) & ($\ast$) \\
    \hline
    SMARTA+ & 92.54 & \textbf{93.30} & 90.96 & 90.77 & 91.63 & 91.91 & 55.58 & 36.14 & 91.44 & 92.19 & \textbf{90.77} & 92.24 \\
    (whole signal) & (0.02) & (0.27) & (0.26) & (1.00) & (0.35) & (0.82) & (0.48) & (0.03) & (0.59) & (0.10) & (0.06) & (0.45) \\
    \hline
    SMARTA+ & \multirow{2}{*}{92.48} & \multirow{2}{*}{93.29} & \multirow{2}{*}{\textbf{91.14}} & \multirow{2}{*}{\textbf{90.80}} & \multirow{2}{*}{\textbf{91.69}} & \multirow{2}{*}{\textbf{91.92}} & \multirow{2}{*}{\textbf{54.67}} & \multirow{2}{*}{\textbf{35.68}} & \multirow{2}{*}{\textbf{91.62}} & \multirow{2}{*}{\textbf{92.25}} & \multirow{2}{*}{90.58} & \multirow{2}{*}{\textbf{92.31}} \\
    (stable periods) & & & & & & & & & & & & \\
    \hline
\end{tabular}
\caption{\label{tab:TPFP} Temporal event localization accuracy comparison of the aDBS algorithm before and after applying artifact removal methods. The p-value obtained using the paired Wilcoxon signed-rank test between SMARTA+ (stable periods) and another algorithm is shown in the parenthesis, where $\ast$ means $p<10^{-3}$.}
\end{table}

\subsection{Low sampling rate}\label{section low sampling rate}
Table~\ref{tab:semiReal_downSamp} presents the results obtained from semi-real signals resampled to 1.1 kHz, aiming to evaluate the performance of SMARTA+ on data with a sampling rate closer to that used in clinical DBS systems. Although the NMSE and SC scores of SMARTA and SMARTA+ showed slight degradation, likely due to distortion introduced during resampling, both methods consistently outperformed template subtraction and blanking. For the AR metric, only three and six points were available for computing $\hat{Z}$ and $\hat{Z}$ in Eq.~\ref{eq:ar}, respectively, which biased the median estimate. Consequently, the first term in Eq.~\ref{eq:ar} became large, suggesting that AR may not be an appropriate measure for evaluating low-sampling-rate signals. Overall, these results indicate that SMARTA+ remains effective even for signals acquired at reduced sampling rates.

Table~\ref{tab:semiReal_downSamp} also summarizes the NMSE across different frequency bands. Note that VHFO components could not be evaluated for signals sampled at 1.1 kHz. Overall, the results were consistent with those in Table~\ref{tab:NMSE}, with the exception that transient blanking achieved the lowest NMSE in the $\gamma$ band. This likely reflects the fact that transient blanking is less affected by resampling-induced distortion compared with template-based methods.

The temporal event localization results for the 1.1 kHz signals are presented in Table~\ref{tab:TPFP_downSamp}. Although overall performance decreased, the findings remained consistent with those in Table~\ref{tab:TPFP}.

\begin{table}[!ht]
\scriptsize
\centering
\begin{tabular}{ccccccccc} 
    \hline
    Median & \multirow{2}{*}{NMSE} & \multirow{2}{*}{AR} & \multirow{2}{*}{SC} & \multirow{2}{*}{$\mathrm{T}_c$ (ms)} & \multirow{2}{*}{$\alpha$} & \multirow{2}{*}{$\beta$} & \multirow{2}{*}{$\gamma$} & \multirow{2}{*}{HFO}\\
    (MAD) & & & & & & & & \\
    \hline
    \multirow{2}{*}{Raw data} & 15.8987 & 1.1925 & 0.4378 & N/A & 9.2494 & -3.4084 & 7.1020 & 22.0712 \\
    & (3.6016) & (0.4462) & (0.0940) & N/A & (3.4453) & (4.4811) & (3.7120) & (3.7425) \\
    \hline
    Template & 11.0212 & 0.7166 & 0.2220 & \textbf{0.0299} & 9.4507 & -3.8814 & 7.7599 & 16.4511 \\
    subtraction & (3.7542) & (0.2589) & (0.0536) & (0.0011) & (3.6498) & (4.6678) & (4.6230) & (4.8424) \\
    \hline
    Pulse & 15.9391 & 0.8938 & 0.3290 & 0.1001 & 9.6025 & -2.5491 & 14.8643 & 21.1013\\
    blanking & (3.6495) & (0.2144) & (0.0345) & (0.0024) & (3.0235) & (4.0273) & (3.6778) & (3.7700) \\
    \hline
    Transient & 7.7286 & \textbf{0.2231} & 0.1634 & 0.1021 & 3.8334 & -3.2304 & \textbf{1.5065} & 5.1021\\
    blanking & (5.1407) & ($<10^{-4}$) & (0.0941) & (0.0032) & (3.3030) & (1.5004) & (5.2869) & (6.8343) \\
    \hline
    \multirow{2}{*}{SMARTA} & 0.5787 & 0.5157 & \textbf{0.1175} & 34.9759 & 9.0866 & -4.1982 & 3.0470 & \textbf{-0.6028} \\
    & (4.1327) & (0.0171) & (0.1175) & (0.2794) & (3.2960) & (3.6839) & (3.2201) & (2.3685)\\
    \hline
    SMARTA+ & -2.2357 & 0.5332 & 0.1808 & 2.9273 & -4.6041 & -10.1709 & 2.9351 & 3.6570 \\
    (whole signal) & (3.4422) & (0.0255) & (0.0462) & (0.1810) & (1.4740) & (4.2665) & (2.8059) & (4.1167) \\
    \hline
    SMARTA+ & \textbf{-3.5628} & 0.5343 & 0.1547 & 2.9042 & \textbf{-5.2826} & \textbf{-10.3744} & 2.1548 & 2.3149 \\
    (stable periods) & (3.0709) & (0.0219) & (0.0447) & (0.1761) & (1.5635) & (4.1279) & (3.0291) & (3.6215) \\
    \hline
    \end{tabular}
\caption{\label{tab:semiReal_downSamp} Performance comparison of artifact removal methods on semi-real aDBS signals. The last four columns show the band-specific NMSE values. All signals were resampled to 1.1 kHz.}
\end{table}

\begin{table}[!ht]
\scriptsize
\centering
\begin{tabular}{cccccccccccccc} 
    \hline
    & \multicolumn{2}{c}{Recall (\%)} & \multicolumn{2}{c}{Precision (\%)} & \multicolumn{2}{c}{F1-score (\%)} & \multicolumn{2}{c}{deviation (ms)} & \multicolumn{2}{c}{OR$_1$ (\%)} & \multicolumn{2}{c}{OR$_2$ (\%)}\\
     & Onset & All & Onset & All & Onset & All & Onset & All & Onset & All & Onset & All\\
    \hline
    \multirow{2}{*}{Raw filtering} & 91.78 & 91.80 & 58.38 & 77.30 & 69.28 & 83.61 & 172.51 & 101.42 & 71.64 & 81.13 & 87.86 & 89.71 \\
    & (0.32) & (0.64) & ($\ast$) & ($\ast$) & ($\ast$) & ($\ast$) & ($\ast$) & ($\ast$) & ($\ast$) & ($\ast$) & (0.08) & (0.02) \\
    \hline
    Template & \textbf{92.50} & 91.84 & 60.21 & 76.99 & 71.09 & 83.44 & 181.43 & 102.67 & 71.14 & 80.22 & 87.69 & 90.71 \\
    subtraction & (0.19) & (0.65) & ($\ast$) & ($\ast$) & ($\ast$) & ($\ast$) & ($\ast$) & ($\ast$) & ($\ast$) & ($\ast$) & (0.07) & (0.69) \\
    \hline
    \multirow{2}{*}{Blanking} & 91.70 & 91.13 & 56.11 & 74.82 & 67.63 & 81.75 & 210.28 & 123.47 & 69.38 & 79.16 & 86.90 & 88.15 \\
    & (0.58) & (0.26) & ($\ast$) & ($\ast$) & ($\ast$) & ($\ast$) & ($\ast$) & ($\ast$) & ($\ast$) & ($\ast$) & (0.02) & ($\ast$) \\
    \hline
    \multirow{2}{*}{SMARTA} & 92.20 & 91.04 & 61.87 & 77.33 & 72.05 & 83.29 & 176.04 & 102.80 & 72.03 & 80.86 & 88.08 & 89.71\\
    & (0.20) & (0.03) & ($\ast$) & ($\ast$) & ($\ast$) & ($\ast$) & ($\ast$) & ($\ast$) & ($\ast$) & ($\ast$) & (0.13) & (0.02) \\
    \hline
    SMARTA+ & 90.45 & 92.15 & 86.13 & 86.42 & \textbf{87.97} & 89.02 & 80.10 & 48.84 & 87.23 & 89.50 & 89.39 & 90.95 \\
    (whole signal) & (0.005) & (0.94) & (0.46) & (0.01) & (0.30) & (0.04) & (0.50) & (0.007) & (0.21) & (0.002) & (0.49) & (0.51) \\
    \hline
    SMARTA+ & \multirow{2}{*}{90.17} & \multirow{2}{*}{\textbf{92.18}} & \multirow{2}{*}{\textbf{86.44}} & \multirow{2}{*}{\textbf{86.63}} & \multirow{2}{*}{87.95} & \multirow{2}{*}{\textbf{89.14}} & \multirow{2}{*}{\textbf{78.64}} & \multirow{2}{*}{\textbf{48.12}} & \multirow{2}{*}{\textbf{87.62}} & \multirow{2}{*}{\textbf{89.63}} & \multirow{2}{*}{\textbf{89.40}} & \multirow{2}{*}{\textbf{91.02}} \\
    (stable periods) & & & & & & & & & & & & \\
    \hline
\end{tabular}
\caption{\label{tab:TPFP_downSamp} Temporal event localization accuracy comparison of the aDBS algorithm before and after applying artifact removal methods. All signals were resampled to 1.1 kHz.}
\end{table}

{
\subsection{Different stimulation frequency}\label{section different stimulation frequency}
To evaluate the flexibility of SMARTA+ under different stimulation settings, semi-real aDBS signals with a stimulation frequency of 180 Hz were generated and used for artifact-removal assessment. The total number of reconstructed artifacts per signal was matched to the 130 Hz condition, while the inter-pulse interval was adjusted to 1/180 s. As a result, the stimulation periods in this dataset were shorter than those in the original aDBS recordings and in the semi-real signals at 130 Hz.
Table~\ref{tab:semiReal_diffRate} presents the performance scores for all methods tested on the 180 Hz signals. The results were consistent with those in Table~\ref{fig:semiReal}, with the exception of the AR value for transient blanking. This difference arises because the blanking duration was fixed at 550 ms; thus, due to the shorter stimulation periods at 180 Hz, most stimulation segments were blanked out. Consequently, the AR index was computed largely over flat segments, resulting in values approaching zero.

The findings for the NMSE scores across frequency bands and the temporal localization results for the 180-Hz signals, which is presented in Table~\ref{tab:TPFP_diffRate}, are consistent with those observed for the 130-Hz semi-real signals.

\begin{table}[!htb]
\scriptsize
\centering
\begin{tabular}{ccccccccccc} 
    \hline
    Median & \multirow{2}{*}{NMSE} & \multirow{2}{*}{AR} & \multirow{2}{*}{SC} & \multirow{2}{*}{$\mathrm{T}_c$ (ms)} & \multirow{2}{*}{$\alpha$} & \multirow{2}{*}{$\beta$} & \multirow{2}{*}{$\gamma$} & \multirow{2}{*}{HFO} & \multirow{2}{*}{VHFO$_1$} & \multirow{2}{*}{VHFO$_2$} \\
    MAD & & & & & & & & & & \\
    \hline
    \multirow{2}{*}{Raw data} & 22.4071 & 2.9329 & 0.1220 & N/A & 9.5436 & -0.9062 & 4.9099 & 12.8157 & 12.4137 & 17.0511\\
    & (5.9845) & (0.1845) & (0.0067) & N/A & (3.9991) & (4.4513) & (3.0578) & (3.3711) & (3.2385) & (3.6469) \\
    \hline
    Template & 17.9437 & 2.9593 & 0.1758 & \textbf{0.0402} & 8.5478 & -2.1402 & 4.7517 & 9.0439 & 9.5489 & 14.8217 \\
    subtraction & (5.8616) & (0.2424) & (0.0185) & (0.0008) & (4.0624) & (4.5610) & (2.7667) & (3.4116) & (3.1729) & (4.0371) \\
    \hline
    Pulse & 9.7193 & 0.3279 & 0.1789 & 0.0916 & 8.9567 & -1.2660 & 6.8580 & 11.6775 & 9.1690 & 11.4547\\
    blanking & (3.4355) & (0.1497) & (0.0422) & (0.0015) & (4.3540) & (4.4017) & (3.8403) & (3.8889) & (3.3216) & (3.8824) \\
    \hline
    Transient & 7.3404 & \textbf{0.0004} & 0.1636 & 0.0979 & 5.5571 & -1.7524 & 0.0221 & 6.9245 & 7.1249 & 8.4059 \\
    blanking & (2.9605) & ($<10^{-4}$) & (0.0257) & (0.0025) & (2.8232) & (2.3255) & (2.9927) & (3.6900) & (3.4426) & (3.9062) \\
    \hline
    \multirow{2}{*}{SMARTA} & 2.7747 & 0.0547 & 0.2061 & 74.8882 & 8.3759 & -1.8588 & 0.3693 & -0.5113 & -9.6180 & \textbf{-11.4659} \\
    & (3.8402) & (0.0047) & (0.0068) & (0.7511) & (4.4234) & (4.8607) & (2.7084) & (3.0941) & (2.3253) & (2.5624) \\
    \hline
    SMARTA+ & -3.2299 & 0.0436 & 0.0592 & 2.0086 & -4.9586 & -10.9213 & -1.9033 & -5.5759 & -7.6421 & -8.1893 \\
    (whole signal) & (4.2170) & (0.0031) & (0.0095) & (0.0760) & (1.3371) & (3.5000) & (3.1899) & (3.6052) & (3.1344) & (2.6886) \\
    \hline
    SMARTA+ & \textbf{-8.5191} & 0.0422 & \textbf{0.0555} & 2.0130 & \textbf{-5.0670} & \textbf{-11.0432} & \textbf{-2.6836} & \textbf{-7.9150} & \textbf{-9.7905} & -10.9005 \\
    (stable periods) & (2.4420) & (0.0027) & (0.0056) & (0.0688) & (1.5979) & (3.7163) & (3.2190) & (2.8310) & (2.4554) & (1.6522) \\
    \hline
    \end{tabular}
\caption{\label{tab:semiReal_diffRate} Performance comparison of artifact removal methods on semi-real aDBS signals with a stimulation frequency of 180 Hz. The last six columns show band-specific NMSE values.}
\end{table}

\begin{table}[!ht]
\scriptsize
\centering
\begin{tabular}{cccccccccccccc} 
    \hline
    & \multicolumn{2}{c}{Recall (\%)} & \multicolumn{2}{c}{Precision (\%)} & \multicolumn{2}{c}{F1-score (\%)} & \multicolumn{2}{c}{deviation (ms)} & \multicolumn{2}{c}{OR$_1$ (\%)} & \multicolumn{2}{c}{OR$_2$ (\%)}\\
     & Onset & All & Onset & All & Onset & All & Onset & All & Onset & All & Onset & All\\
    \hline
    \multirow{2}{*}{Raw filtering} & 93.24 & 91.46 & 63.09 & 77.58 & 73.01 & 83.51 & 214.21 & 109.35 & 71.98 & 82.17 & 87.06 & 91.15 \\
    & (0.39) & (0.005) & ($\ast$) & ($\ast$) & ($\ast$) & ($\ast$) & ($\ast$) & ($\ast$) & ($\ast$) & ($\ast$) & (0.003) & (0.13) \\
    \hline
    Template & 92.85 & 92.13 & 63.89 & 77.42 & 73.35 & 83.68 & 202.08 & 104.07 & 72.68 & 82.18 & 86.91 & 90.63 \\
    subtraction & (0.19) & (0.03) & ($\ast$) & ($\ast$) & ($\ast$) & ($\ast$) & ($\ast$) & ($\ast$) & ($\ast$) & ($\ast$) & ($\ast$) & (0.03) \\
    \hline
    \multirow{2}{*}{Blanking} & 93.76 & 91.88 & 69.27 & 78.46 & 77.73 & 84.10 & 196.61 & 103.20 & 74.18 & 82.62 & 87.66 & 90.96 \\
    & (0.50) & (0.03) & ($\ast$) & ($\ast$) & ($\ast$) & ($\ast$) & ($\ast$) & ($\ast$) & ($\ast$) & ($\ast$) & (0.002) & (0.06) \\
    \hline
    \multirow{2}{*}{SMARTA} & 92.77 & 92.25 & 66.32 & 78.39 & 75.18 & 84.39 & 183.48 & 99.04 & 74.53 & 82.70 & 88.22 & 90.68\\
    & (0.18) & (0.07) & ($\ast$) & ($\ast$) & ($\ast$) & ($\ast$) & ($\ast$) & ($\ast$) & ($\ast$) & ($\ast$) & (0.01) & (0.009) \\
    \hline
    SMARTA+ & 94.38 & \textbf{93.63} & 91.84 & 90.29 & 92.99 & 91.86 & 61.44 & 38.68 & 91.28 & 92.45 & 90.72 & 92.01 \\
    (whole signal) & (0.31) & (0.20) & (0.91) & (0.60) & (0.52) & (0.60) & (0.17) & ($\ast$) & (0.63) & (0.005) & (0.63) & (0.06) \\
    \hline
    SMARTA+ & \multirow{2}{*}{\textbf{94.58}} & \multirow{2}{*}{93.62} & \multirow{2}{*}{\textbf{92.40}} & \multirow{2}{*}{\textbf{90.30}} & \multirow{2}{*}{\textbf{93.35}} & \multirow{2}{*}{\textbf{91.87}} & \multirow{2}{*}{\textbf{60.29}} & \multirow{2}{*}{\textbf{37.91}} & \multirow{2}{*}{\textbf{91.39}} & \multirow{2}{*}{\textbf{92.60}} & \multirow{2}{*}{\textbf{90.82}} & \multirow{2}{*}{\textbf{92.17}} \\
    (stable periods) & & & & & & & & & & & &  \\
    \hline
\end{tabular}
\caption{\label{tab:TPFP_diffRate} Temporal event localization accuracy comparison of the aDBS algorithm before and after applying artifact removal methods on semi-real signals with a stimulation frequency of 180 Hz.}
\end{table}

}

\section{Discussion}
 
Stimulation-induced artifacts, including stimulus and transient DC components, pose major challenges for aDBS implementation. Effective artifact removal is critical for accurate online neural signal detection and reliable feedback control. We proposed SMARTA+, a back-end enhancement of SMARTA \cite{liu2024}, aimed at improving computational efficiency for real-time implementation and transient DC artifact suppression while maintaining artifact-removal performance. Key improvements include replacing the KNN search with ANN and incorporating wavelet-based dimension reduction, reducing computation time to under 10\% of that required by SMARTA. SMARTA+ achieved comparable artifact suppression (as measured by AR and SC) and demonstrated improved recovery of clean signals, reflected in lower NMSE over different spectral bands. The temporal event localization analysis further confirmed its effectiveness in accurately detecting beta-burst events. To our knowledge, this is the first study to evaluate how accurately a stimulus artifact removal algorithm can enhance beta burst detection.
 
We now discuss why SMARTA+ achieves the encouraging NMSE results. Although the $\mathrm{NMSE}_\beta$ for the semi-real aDBS signals was as low as -3.4, as shown in Table \ref{tab:NMSE}, reflecting the benefits of carefully chosen sampling rates and stimulation frequencies, the high recall but low precision in beta burst detection from raw signals indicates limited effectiveness. Stimulation artifacts might still mislead aDBS algorithms, causing frequent false detections despite the optimized signal design for PD therapy.
This issue arises mainly from two factors. First, while the sampling rate and stimulation frequency have been carefully designed to avoid aliasing, the time-varying morphology of stimulus artifacts still introduces spectral leakage via aliasing. To appreciate this statement, consider the following simple model. Suppose the stimulus artifacts are modeled as $f(t)=A(t)\sum_{j\in \mathbb{Z}}s(t-t_j)$, where $s$ is smooth and compactly supported modeling the morphology of stimulus artifact so that $s(t)=\alpha_0+\sum_{k=1}^\infty \alpha_k \cos(2\pi kt+\beta_k)$, $\alpha_0\in \mathbb{R}$ is the mean value of $s$, $\alpha_k\geq 0$ and $\beta_k\in [0,2\pi)$, $t_j=j/130$ is the timing of stimulation and $A(t)$ is a positive smooth function modeling the strength of stimulus artifact. As discussed in \cite{liu2024}, due to the soft-matter properties of the brain, the stimulus artifact generally changes from time to time, particularly the strength. Despite its simple behavior, we know that $\hat{f}(\xi)=\alpha_0\hat{A}(\xi)+\frac{1}{2}\sum_{k=1}^\infty \alpha_k [e^{\beta_k}\hat{A}(\xi-130k) + e^{-\beta_k}\hat{A}(\xi+130k)]$, where the equality holds uniformly. Therefore, when $A$ is large, unless $A$ is constant or the support of $\hat{A}$ is compact and ``very'' concentrated around $0$, the spectral spreading of $\hat{A}$ could contaminate the beta band. 
Second, abrupt stimulus onsets and transient DC artifacts introduce broadband distortions that overlap with true beta activity, inflating beta amplitude and triggering false detections. This can be specifically seen in the improvement of NMSE${}_\alpha$ shown in Table \ref{tab:NMSE}.
Artifact removal using template subtraction, pulse blanking, or SMARTA improves precision and F1-score by reducing these false positives. However, residual transient artifacts, especially at stimulation onset, remain challenging. SMARTA+ addresses this by also targeting onset-related DC transients, resulting in further gains in precision and F1-score, especially during stimulation onset.

A directly related topic is the existing aDBS implementations. On the Summit RC+S system (Medtronic), aDBS is challenged by both stimulus artifacts and DC transients \cite{anso2022artifactRemoval}. While front-end blanking effectively removes stimulus artifacts, residual DC transients introduce broadband noise that can degrade beta burst detection and lead to erroneous self-triggering. In \cite{anso2022artifactRemoval}, this was mitigated by back-end detector blanking with a 550 ms window. However, such a long blanking duration risks omitting relevant beta activity shortly after stimulation, potentially impairing aDBS performance.
An alternative approach was explored in \cite{thenaisie2021percept}, where LFPs recorded using the Percept PC system were analyzed. By carefully aligning the stimulation frequency with the sampling rate, stimulus artifacts were minimized within the beta band. However, the study \cite{thenaisie2021percept}  did not implement aDBS or address DC transient artifacts, leaving it unclear whether self-triggering would still occur under aDBS deployment. Also, the spectral leakage via aliasing due to time-varying stimulus artifact morphology is also inevitable in this situation.

As a back-end method, SMARTA+ is adaptable to various neuromodulation platforms, including Summit RC+S and Percept PC, for real-time removal of both stimulus and transient artifacts. 
The initial decision trees used by SMARTA+ can be pre-built, with updates during stimulation-off periods when the intensity ramps down and the aDBS controller continues collecting samples for online beta-amplitude estimation \cite{little2013aDBS,tinkhauser2017aDBS}. As shown in Tables~\ref{tab:semiReal} and \ref{tab:real}, SMARTA+ processes each artifact segment faster than the stimulation interval at a 130-Hz sampling rate. These results indicate that SMARTA+ is well suited for real-time aDBS deployment, where its ability to suppress artifacts at each stimulation onset enables more reliable beta detection and may enhance adaptive stimulation control.

The amplitude of transient artifacts is known to correlate with the rate of change in stimulation intensity \cite{hammer2022deltaStimulation}. A common mitigation strategy is to modulate the stimulation between predefined upper and lower bounds instead of switching it fully on or off \cite{velisar2019aDBS}. This controlled ramping reduces the rate of change and thereby attenuates transient artifacts. Nevertheless, despite the reduction in artifact severity, both transient DC artifacts and stimulus artifacts are not entirely eliminated, and the degree of reduction depends critically on the chosen lower amplitude limit. In certain cases, this lower bound may approach zero, yielding substantial artifacts despite dual-threshold control. Thus, while dual-threshold aDBS helps reduce artifact severity, applying SMARTA+ as an additional processing step remains valuable for further improving signal quality.

Artifact contamination also constrains stimulation parameters in aDBS. For example, the above-mentioned frequency planning is often required to prevent aliasing peaks from overlapping with the beta band \cite{thenaisie2021percept}. Although cDBS allows broader frequency tuning, aDBS is more restricted unless artifacts are effectively suppressed. By reliably removing stimulus artifacts, SMARTA+ relaxes these limitations and enables more flexible frequency selection.

Electrode configuration presents another challenge. Common-mode rejection typically requires placing the stimulation contact between two recording contacts, restricting viable electrode arrangements in clinical practice. Although directional electrodes offer improved targeting \cite{tinkhauser2018directional}, their use in aDBS has been limited by the need for common-mode rejection. While SMARTA+ was demonstrated here using bipolar recordings, it is also effective for unipolar signals, thereby facilitating the integration of directional electrodes into aDBS systems while maintaining robust artifact suppression.

Beta-band LFP power is the most widely used control biomarker for aDBS in Parkinson’s disease, yet beta activity alone may not fully represent the underlying neural state \cite{johnson2016beyondbeta}. Gamma-band activity correlates with movement and motor performance \cite{swann2018aDBSgamma,lofredi2018aDBSgamma}, and alpha band dynamics are linked to cognitive and emotional processing \cite{marceglia2011alpha}, motivating the use of multi-band biomarkers. However, incorporating multiple frequency features becomes difficult when stimulation artifacts are not adequately addressed, as stimulation frequencies must then be chosen to avoid aliasing across all bands of interest. By suppressing artifacts across the entire spectrum, SMARTA+ enables more accurate recovery of LFP activity, supports the use of diverse biomarkers, and paves the way for more flexible and precise aDBS strategies, an avenue we will explore in future work. 

Adaptive DBS is also being investigated for disorders beyond PD, using control biomarkers derived from signals other than subthalamic LFPs. In essential tremor, for example, low-frequency oscillations recorded from subdural cortical electrodes over the primary motor cortex and from the ventral intermediate nucleus have been used for aDBS control \cite{opri2020ET}. While clinical aDBS trials for cognitive or psychiatric applications remain limited, neurophysiological studies indicate strong potential \cite{guidetti2021aDBSintro}. Such applications may require diverse biomarkers across multiple brain regions and recording modalities. A key advantage of SMARTA+ is its versatility. Theoretically, it is not restricted to LFPs and can be adapted to different signal types, sampling rates, and stimulation frequencies, as partially demonstrated in Sections \ref{section low sampling rate} and \ref{section different stimulation frequency}. This flexibility positions SMARTA+ as a promising artifact-removal tool across a broad range of aDBS research settings, supporting the development of generalizable and disorder-specific adaptive stimulation systems. For example, VHFOs ($>$ 1000 Hz) in EEG signals are relevant to epilepsy \cite{usui2015VHFO} and can serve as potential biomarkers. However, these VHFOs can be contaminated by stimulus artifacts if not properly processed. Applying SMARTA+ to such signals may facilitate artifact-free VHFO extraction and contribute to advancing aDBS approaches for epilepsy.

This work has several limitations. The dedicated hardware required to implement the front-end approaches described in \cite{culaclii2016ts,stanslaski2018aDBSdevice,anso2022artifactRemoval} was not available in this study; therefore, these methods were simulated using software-based implementations. As a result, their performance may not fully reflect the optimal capabilities of the original hardware systems. Nevertheless, SMARTA+ was evaluated under identical conditions, without incorporating any front-end artifact removal processes. While SMARTA+ is substantially faster than SMARTA, it remains computationally intensive compared with traditional back-end template subtraction or blanking algorithms, particularly when considering battery constraints. Addressing this will require not only advances in energy-efficient hardware but also further algorithmic optimization. Future directions include incorporating randomized algorithms \cite{jones2011RANN,jones2013RANN} and exploring cloud-based processing. As an algorithm-focused study, we did not explore leveraging hardware properties to further optimize SMARTA+. Such considerations are essential for clinical implementation but lie beyond the current scope and will be addressed in future work. The study population is relatively small and limited to Asian subjects from a single hospital. Larger-scale studies are needed to further validate the proposed algorithm.

\section{Conclusion}
The SMARTA+ algorithm is proposed as an enhanced and time-efficient extension of SMARTA, with demonstrated effectiveness in removing artifacts in LFP signals recorded during electrical stimulation. For each artifact, an accurate template is constructed by identifying similar segments from both the target signal and a pre-recorded dataset using an approximate nearest neighbors algorithm, guided by a similarity metric based on optimal shrinkage for noise reduction. The underlying LFP signal is then recovered by subtracting this template. Additionally, DC transient artifacts are estimated through a two-stage process involving signal smoothing and polynomial fitting and are subtracted from the signal. Temporal event localization analysis shows that SMARTA+ improves the precision of beta-event detection and reduces the timing deviation of detected events. With a computational cost lower than the duration of each artifact segment, SMARTA+ demonstrates suitability for integration into real-time aDBS algorithms for artifact removal.

\section{Ethics statement}
This study was reviewed and approved by the Chang Gung Medical Foundation Institutional Review Board (CGMH-IRB No. 202500934B0) for secondary analysis of de-identified data obtained from prior IRB-approved studies. Informed consent had been obtained from all participants under the original IRB protocols. The research was conducted in accordance with the principles of the Declaration of Helsinki.

\section{Funding}
This work was supported by the National Science and Technology Council, Taiwan (NSTC112-2314-B-182-042-MY3, NSTC112-2115-M-002-015-MY3, NSTC114-2811-M-002-158) and the National Health Research Institutes, Taiwan (NHRI-EX113-11104NI).

\section*{Reference}
\bibliographystyle{abbrv}
\bibliography{SMARTAbib}

\clearpage
\appendix
\section{eOptShrink} \label{sec:cOS}
After segmentation, the LFPs in the matrix $\XX$ were removed using the extended optimal shrinkage (eOptShrink) algorithm \cite{su2022cOS}. For the sake of completeness, we summarize eOptShrink here. Without loss of generality, assume $p\leq n$ and denote $\beta=p/n$. Denote the singular value decomposition (SVD) of $\XX$ as

\begin{equation}
    \XX=\sum^{p}_{i=1}\tilde{\sigma}_i \tilde{\boldsymbol u}_i \tilde{\boldsymbol v}_i^\top,
\end{equation}
where $\tilde{\sigma}_i$ is the $i$th singular value, satisfying $\tilde{\sigma}_1\geq \tilde{\sigma}_2\geq \ldots \geq\tilde{\sigma}_p\geq 0$, and $\tilde{\boldsymbol u}_i$ and $\tilde{\boldsymbol v}_i$ are the corresponding left and right singular vectors, respectively, with $\|\tilde{\boldsymbol u}_i\|_2=\|\tilde{\boldsymbol v}_i\|_2=1$. Next, the threshold $\hat{\lambda}_+$ was computed as

\begin{equation} \label{equation lambda+}
\hat{\lambda}_+:=\tilde{\lambda}_{[n^{1/4}]+1}+\frac{1}{2^{2/3}-1}(\tilde{\lambda}_{[n^{1/4}]+1}-\tilde{\lambda}_{2[n^{1/4}]+1}),
\end{equation}
where $[n^{1/4}]$ denotes the closest integer to $n^{1/4}$, and $\tilde{\lambda}_i=\tilde{\sigma}_i^2$ is the $i$th eigenvalue of $\XX\XX^\top$. 

For $i=1,\ldots,\hat{r}$, where $\hat{r}$ is the estimated effective rank given by the number of $\tilde{\lambda_i}$ greater than $\hat{\lambda}_++n^{-1/3}$; that is,

\begin{equation} \label{equation rhat}
    \hat{r}=\left|\left\{\tilde{\lambda}_i|\tilde{\lambda}_i>\hat{\lambda}_++n^{-1/3}\right\}\right|,
\end{equation}
the following value was computed:

\begin{equation} \label{equation di'}
    \hat{d}_i=\hat{\varphi}_i\sqrt{\hat{a}_{1,i}\hat{a}_{2,i}},
\end{equation}
where 

\begin{equation} \label{equation varphii'}
    \hat{\varphi}_i=\sqrt{\frac{1}{\hat{T}_i}},\;\hat{a}_{1,i}=\frac{\hat{m}_{1,i}}{\hat{\varphi}_i^2\hat{T}_i^{'}},\;\hat{a}_{2,i}=\frac{\hat{m}_{2,i}}{\hat{\varphi}_i^2\hat{T}_i^{'}},
\end{equation}

\begin{equation} \label{equation Ti}
    \hat{T}_i=\tilde{\lambda}_i\hat{m}_{1,i}\hat{m}_{2,i},\;\hat{T}_i^{'}=\hat{m}_{1,i}\hat{m}_{2,i}+\tilde{\lambda}_i\hat{m}_{1,i}^{'}\hat{m}_{2,i}+\tilde{\lambda}_i\hat{m}_{1,i}\hat{m}_{2,i}^{'},
\end{equation}

\begin{equation} \label{equation mhat}
    \hat{m}_{1,i}:=\frac{1}{p}\left(\sum_{j=1}^k\frac{1}{\hat{\lambda}_j-\tilde{\lambda}_i}+\sum_{j=k+1}^p\frac{1}{\tilde{\lambda}_j-\tilde{\lambda}_i}\right),\; \hat{m}_{2,i}:=\frac{1-\beta}{\tilde{\lambda}_i}+\beta\hat{m}_{1,i},
\end{equation} 

\begin{equation}\label{equation mhat'}
    \hat{m}_{1,i}^{'}:=\frac{1}{p}\left(\sum_{j=1}^k\frac{1}{(\hat{\lambda}_j-\tilde{\lambda}_i)^2}+\sum_{j=k+1}^p\frac{1}{(\tilde{\lambda}_j-\tilde{\lambda}_i)^2}\right),\;\hat{m}_{2,i}^{'}:=\frac{1-\beta}{\tilde{\lambda}_i^2}+\beta\hat{m}_{1,i},
\end{equation}
and $k<p$ is a given parameter ($k=10$ in this study). Additionally,

\begin{equation} \label{equation lambdaj} 
    \hat{\lambda}_j:=\tilde{\lambda}_{k+1}+\frac{1-(\frac{j-1}{k})^{2/3}}{2^{2/3}-1}(\tilde{\lambda}_{2k+1}-\tilde{\lambda}_k+1).
\end{equation}
Finally, the LFP-free matrix was obtained as

\begin{equation}\label{definition Shat}
    \hat{\SX}=\sum_{i=1}^{\hat{r}}\hat{d}_i\tilde{\boldsymbol u}_i\tilde{\boldsymbol v}_i^\top=\begin{bmatrix} \hat{\boldsymbol{s}}_1 & \hat{\boldsymbol{s}}_2 & \ldots & \hat{\boldsymbol{s}}_n \end{bmatrix}.
\end{equation}
To accelerate the eOptShrink algorithm, the randomization technique described in \cite{halko2011finding} was applied. %For implementation details, see \cite{liu2024}.
\end{document}